\documentclass[twocolumn,showpacs,amsmath,pra,amssymb,superscriptaddress]{revtex4-1}

\pdfoutput=1

\usepackage[export]{adjustbox}
\usepackage{amsfonts}
\usepackage{amsmath}
\usepackage[makeroom]{cancel}
\usepackage[english]{babel}
\usepackage[T1]{fontenc}
\usepackage{times}
\usepackage{mathrsfs}
\usepackage{graphicx}
\usepackage{dcolumn}
\usepackage{bm}
\usepackage{wasysym}
\usepackage[colorlinks,bookmarks=true,citecolor=blue,linkcolor=red,urlcolor=blue]{hyperref}
\usepackage[tight, FIGTOPCAP, hang, raggedright, nooneline]{subfigure}
\usepackage{hyperref}
\usepackage{xcolor}
\usepackage{epsfig}
\usepackage{amssymb}
\usepackage{lipsum}
\usepackage{algorithm}
\usepackage{algpseudocode}

\makeatletter
\def\p@subsection{}
\makeatother

\newcommand{\subfigimg}[3][,]{%
  \setbox1=\hbox{\includegraphics[#1]{#3}}
  \leavevmode\rlap{\usebox1}
  \rlap{\hspace*{0pt}\raisebox{\dimexpr\ht1+0\baselineskip}{#2}}
  \phantom{\usebox1}
  }
  
 \definecolor{Green}{RGB}{80,182,0}

\usepackage{color}

\newcommand{\la}{\langle}
\newcommand{\ra}{\rangle}

\graphicspath{./Images/}
\usepackage{epstopdf}

\begin{document}
\title{Dynamical Localization in $\mathbb{Z}_2$ Lattice Gauge Theories}
\author{Adam Smith}
\email{as2457@cam.ac.uk}
\affiliation{T.C.M. group, Cavendish Laboratory, J.~J.~Thomson Avenue, Cambridge, CB3 0HE, United Kingdom}
\author{Johannes Knolle}
\affiliation{T.C.M. group, Cavendish Laboratory, J.~J.~Thomson Avenue, Cambridge, CB3 0HE, United Kingdom}
\affiliation{Blackett Laboratory, Imperial College London, London SW7 2AZ, United Kingdom}
\author{Roderich Moessner}
\affiliation{Max Planck Institute for the Physics of Complex Systems, N\"{o}thnitzer Stra{\ss}e 38, 01187 Dresden, Germany}
\author{Dmitry L.~Kovrizhin}
\affiliation{Rudolf Peierls Centre for Theoretical Physics, 1 Keble Road, Oxford, OX1 3NP, United Kingdom}
\affiliation{NRC Kurchatov institute, 1 Kurchatov square, 123182, Moscow, Russia}
\date{\today}

\begin{abstract}
We study quantum quenches in two-dimensional lattice gauge theories with fermions coupled to dynamical $\mathbb{Z}_2$ gauge fields. Through the identification of an extensive set of conserved quantities, we propose a generic mechanism of charge localization in the absence of quenched disorder both in the Hamiltonian and in the initial states. We provide diagnostics of this localization through a set of experimentally relevant dynamical measures, entanglement measures, as well as spectral properties of the model. One of the defining features of the models that we study is a binary nature of emergent disorder, related to $\mathbb{Z}_2$ degrees of freedom. This results in a qualitatively different behaviour in the strong disorder limit compared to  typically studied models of localization. For example it gives rise to a possibility of a delocalization transition via a mechanism of quantum percolation in dimensions higher than 1D. We highlight the importance of our general phenomenology to questions related to dynamics of defects in Kitaev's toric code, and to quantum quenches in Hubbard models. While the simplest models we consider are effectively non-interacting, we also include interactions leading to many-body localization-like logarithmic entanglement growth. Finally, we consider effects of interactions that generate dynamics for conserved charges, which gives rise to only transient localization behaviour, or quasi-many-body-localization.
\end{abstract}

\maketitle

\section*{Introduction}

\begin{figure*}[th!]
	\centering
	\includegraphics[width=.9\textwidth]{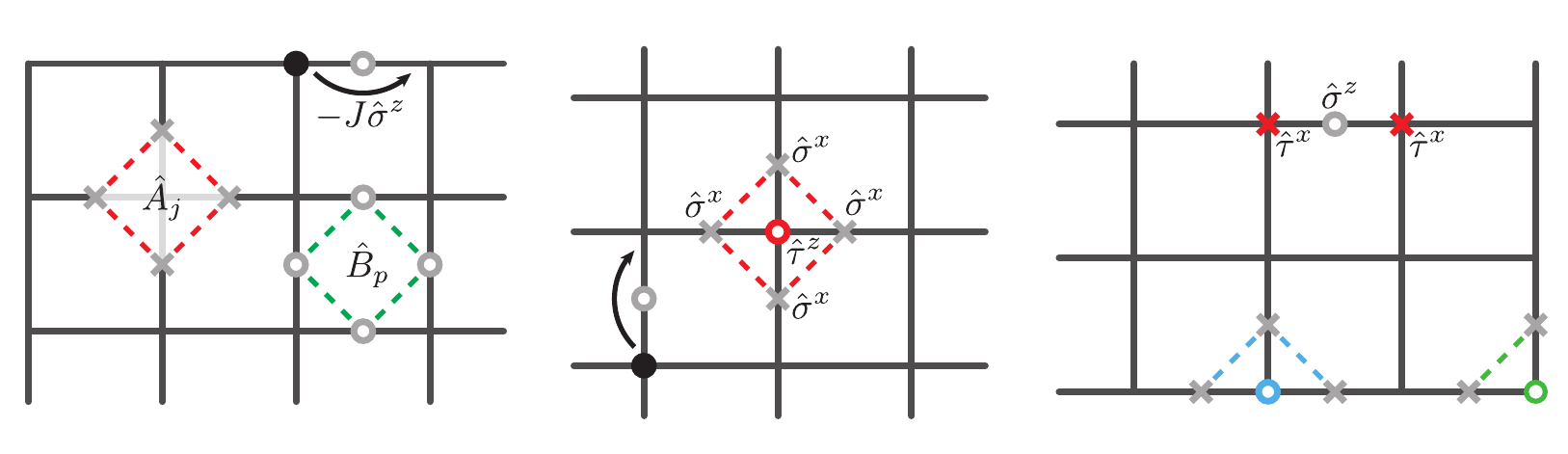}
	\caption{Schematic picture of the model~\eqref{eq: H general}. Left panel shows star $\hat{A}$ and plaquette $\hat{B}$ operators, with $\hat{\sigma}^x$, $\hat{\sigma}^z$ operators denoted by crosses and circles respectively. Fermion hopping is defined by $J$ and the direction of the spin-1/2 on that bond. Centre and right panels show the duality transformation to new spins $\hat{\tau}$. The model can be defined with periodic boundary conditions, as in the centre panel. In the case of open boundaries, we define incomplete boundary ``stars'' as shown in the right panel.}\label{fig: 2D model}
\end{figure*}

Gauge theories play a central role in theoretical physics, most famously in the unified description of fundamental particles in the standard model, but also increasingly in the description of condensed matter systems~\cite{Kogut1979,Fradkin2013,WenBook,Wiese2013,Henley2010}, where lattice gauge theory (LGT) models often arise as effective descriptions of strongly-correlated systems. The celebrated toric code~\cite{KitaevTQM} is such an example of a $\mathbb{Z}_2$ lattice gauge theory which is a prototypical quantum stabiliser code, which also serves as an effective description of the Kitaev honeycomb model with strongly anisotropic couplings~\cite{Kitaev}.

The honeycomb Kitaev model can itself be understood in terms of itinerant Majorana fermions coupled to static $\mathbb{Z}_2$ gauge fields. Other examples include the resonating valence-bond liquid~\cite{Moessner2001,Moessner2001b}, slave-particle descriptions of the Hubbard model~\cite{Ruegg2010,Zitko2015}, non-Fermi metals~\cite{Nandkishore2012} and glasses~\cite{Parameswaran2017}, the Falicov-Kimball model~\cite{Antipov2016,Hermann2017,Gazit2016,Schubert2008}, etc. While models of lattice gauge theories are often difficult to realize in experiment, recent developments in cold atom quantum simulators have opened possibilities in studying these models, see, e.g., the pioneering experiment on cold ion simulations of the Schwinger model~\cite{Martinez2016}. This progress motivates the importance in understanding simple exactly solvable models, which one could use to benchmark experiments and to improve our theoretical understanding of universal behaviour of LGTs. 

Another field of importance in condensed matter physics is localization (Anderson localization~\cite{Anderson1958}, and many-body localization), which has seen recent remarkable developments in theory~\cite{Nandkishore2015,Abanin2017} and experiments~\cite{Choi2016,Schreiber2015}. Localization phenomena provide a set of fundamental concepts about the insulating behaviour of itinerant degrees of freedom in the presence of disorder. Remarkably, localization was shown to persist even in presence of interactions and the resulting many-body localized (MBL) phase is a novel exotic dynamical phase of matter~\cite{Basko2006,Znidaric2008,Bardason2012} which is robust to generic perturbations. Many-body-localization provides a mechanism for non-trivial relaxation, beyond integrable models, and serves as a counter example to eigenstate thermalization~\cite{Srednicki1994}.

Following original ideas of Kagan and Maksimov~\cite{Kagan}, a number of models for disorder-free localization featuring heavy and light particles have been proposed. In this setup, localization may be induced purely via interactions between the two species~\cite{Schiulaz2014,Schiulaz2015,Yao2014,Papic2015} without any quenched disorder in the Hamiltonian. While numerics suggests localization behaviour in these systems, so far, it has been found to be only transient, giving way to ergodic behaviour in the long-time limit. This behaviour was therefore dubbed quasi-MBL~\cite{Yao2014}. Another interesting approach is to take quantum analogues of classically-glassy systems, where non-ergodic behaviour~\cite{Horssen2015,Hickey2016,Garrahan2017} has also been observed. Unfortunately, due to small available system sizes, glassy behaviour in these models has not yet been distinguished from that of quasi-MBL. 

In previous work~\cite{Smith2017,Smith2017_2}, we have demonstrated, for the first time, a general mechanism for disorder-free localization in a model of fermions with a local $\mathbb{Z}_2$ gauge symmetry. We showed how the localization signatures can be revealed through an exact identification between conserved charges associated with a $\mathbb{Z}_2$ gauge symmetry and an effective binary potential for non-interacting fermions. Moreover, we were able to map fermion or spin correlators to disorder averaged fermionic correlators despite both the Hamiltonian and the initial state having no quenched disorder at all.

In this paper, we extend our theory to a family of $\mathbb{Z}_2$ lattice gauge models of spinless fermions coupled to spins-1/2. These models can be defined in any dimension, and for lattices described by arbitrary graphs, and in particular we focus on two-dimensional square lattices. Here we show that the localization mechanism we discussed in the 1D case applies in a more general context. We use experimentally relevant dynamic probes to diagnose  localization~\cite{Choi2016,Schreiber2015}. In two dimensions we are able to make a direct connection between our model and Kitaev's toric code model in presence of dynamical charges.

We also analyse in more detail the limit of strong disorder. In the case of $\mathbb{Z}_2$ gauge degrees of freedom, this disorder takes binary values, which gives rise to a phenomenology not found in the typically studied continuous quenched disorder realizations~\cite{Janarek2017}. We find a mechanism for delocalization in 2D related to the phenomenology of quantum percolation~\cite{Alvermann2005,Schubert2009}. Further, we study perturbations that render our models fully interacting. As our preliminary studies in Ref.~\cite{Smith2017_2} have indicated, here we demonstrate that in presence of perturbations that do not induce dynamics of conserved charges the entanglement entropy is characterised by a logarithmic growth, which can be likened to MBL behaviour. In addition, we provide an analysis of the effects of perturbations that generate dynamics of conserved charges, which leads to quasi-MBL behaviour in the region of parameters that we have explored.

The structure of the paper is the following. In Section~\ref{sec: Model}, we define the family of models in arbitrary spatial dimension and on arbitrary lattices. We identify conserved charges which reveal the general disorder-free mechanism for localization. We  provide details about the transformations, the initial states, and the calculations. We explain the phenomenology of disorder-free localization for the example of 2D Kitaev's toric code and the dynamics of defects therein in Section~\ref{sec: toric code}. In Section~\ref{sec: localization} we present a discussion on the diagnostics of localization behaviour. In Sections~\ref{sec: strong disorder} and~\ref{sec: delocalization}, we focus on the binary nature of the effective disorder, which leads to qualitative differences in behaviour. In Section~\ref{sec: interactions} we discuss the effects of integrability-breaking perturbations. In Section~\ref{sec: conserved} we discuss MBL physics which arises in presence of perturbations, while in Section~\ref{sec: quasi-MBL} we consider terms that give dynamics to our effective disorder, leading to quasi-MBL behaviour. A general discussion and conclusions are presented in Section~\ref{sec: discussion}. We provide in-depth details of all of the numerical methods used in the paper in Appendices.

\section{Model}\label{sec: Model}

We study a family of lattice models with spinless fermions, $\hat{f}_i$, which live on the sites of a lattice minimally coupled to spins-1/2, $\hat{\sigma}_{jk}$, positioned on the bonds. These models can be defined on an arbitrary graph, however, in this paper we focus on one-dimensional chains and a two-dimensional square lattice with both open and periodic boundary conditions. We also discuss three dimensional generalizations, as well as effects of perturbations. The models are described by the Hamiltonian
\begin{equation}\label{eq: H general}
\hat{H} = -  \sum_{\la j k \ra} J_{jk}\hat{\sigma}^z_{jk} \hat{f}^\dagger_j \hat{f}_k - \sum_j h_j\hat{A}_j,
\end{equation}
where $\la j k \ra$ denotes nearest neighbours, and $\hat{A}_j$ is the \emph{star} operator, which is the product of all spins on the bonds connected to site $j$, shown for a 2D square lattice in Fig.~\ref{fig: 2D model},
\begin{equation}
\hat{A}_j = \prod_{k : \la j k \ra} \hat{\sigma}^x_{jk},
\end{equation}
and $J_{jk}$, $h_j$ define coupling strength, and local magnetic field, respectively. In the following we assume that both of them are position independent.
The Hamiltonian posseses an extensive number of conserved quantities (charges) $\hat{q}_i = (-1)^{\hat{n}_i} \hat{A}_i$, where $\hat{n}_i=\hat{f}^{\dagger}_i\hat{f}_i$. The charges have eigenvalues $\pm 1$ and commute with the Hamiltonian and amongst themselves $[\hat{H},\hat{q}_i]=0$, and $[\hat{q}_i,\hat{q}_{j}] = 0$. They can be used to generate local $\mathbb{Z}_2$ gauge transformations under which the Hamiltonian is invariant. Explicitly, these transformations are given by the unitary operators $\hat{U}(\{\theta_i\}) = \prod_i \hat{q}_i^{(1-\theta_i)/2}$, where $\theta_i = \pm 1$, which transform the operators accordingly 
\begin{equation}
\hat{f} \rightarrow \theta_i \hat{f}, \qquad \hat{\sigma}^z_{ij} \rightarrow \theta_i \theta_j \hat{\sigma}^z_{ij}.
\end{equation}
It is worth noting that our model is an example of an unconstrained $\mathbb{Z}_2$ lattice gauge theory. Explicitly while the Hamiltonian is invariant under the gauge transformation, the Hilbert space is not. What is typically understood as a gauge theory is constrained to the physical subspace of gauge invariant states by Gauss law~\cite{Kogut1979,Fradkin2013,WenBook} $\hat{q}_i |\Psi\ra = |\Psi\ra$, which we do not impose in our case, c.f.~the gauge structure of Kitaev honeycomb model~\cite{Kitaev}.

In our previous work~\cite{Smith2017,Smith2017_2} we studied the Hamiltonian~\eqref{eq: H general} defined on a 1D chain. In this case the star operators reduce to nearest-neighbour exchange couplings $\hat{A}_j = \hat{\sigma}^x_{j-1,j} \hat{\sigma}^x_{j,j+1}$, and the Hamiltonian Eq.~\eqref{eq: H general} assumes the following form
\begin{equation}
\hat{H}_{1D} = - J \sum_{\la j k \ra} \hat{\sigma}^z_{jk} \hat{f}^\dagger_j \hat{f}_k - h \sum_j \hat{\sigma}^x_{j-1,j} \hat{\sigma}^x_{j,j+1}.
\end{equation}
Previously, we have shown for this 1D model that conserved charges $\hat{q}_i$ play a role of emergent binary disorder, which gives rise to localization of electron degrees of freedom.

Models described by the class of Hamiltonians~\eqref{eq: H general} appear in the studies of a wide range of systems. Specifically, by imposing a global constraint on the conserved charges such that $\sum_i (\hat{q}_i + 1) |\Psi\ra = \mu |\Psi\ra$, where $\mu\in\mathbb{Z}$, we recover the Falicov-Kimball model~\cite{Antipov2016,Hermann2017,Gazit2016} where $\mu/N$ is the density of localized electrons, see discussion below. Further, in the case of spin-1/2 fermions with the kinetic term described by $\sum_{\alpha=\uparrow,\downarrow,\langle ij\rangle} \hat{\sigma}^z_{ij} \hat{f}^\dag_{i\alpha} \hat{f}_{j\alpha}$ and imposing the Gauss law constraint, we arrive at the slave-boson description of the Hubbard model~\cite{Zitko2015}. For a recent exposition of other interesting directions where the phenomenology of disorder-free localization described in this paper applies see e.g.~Ref.~\cite{Prosko2017}. Also see Ref.~\cite{Brenes2018} which studies a mechanism similar to ours in the Schwinger model with discrete $\mathbb{Z}_2$ symmetry replaced by a continuous $U(1)$ gauge field.

\begin{figure*}[th!]
	\centering
	\includegraphics[width=\textwidth]{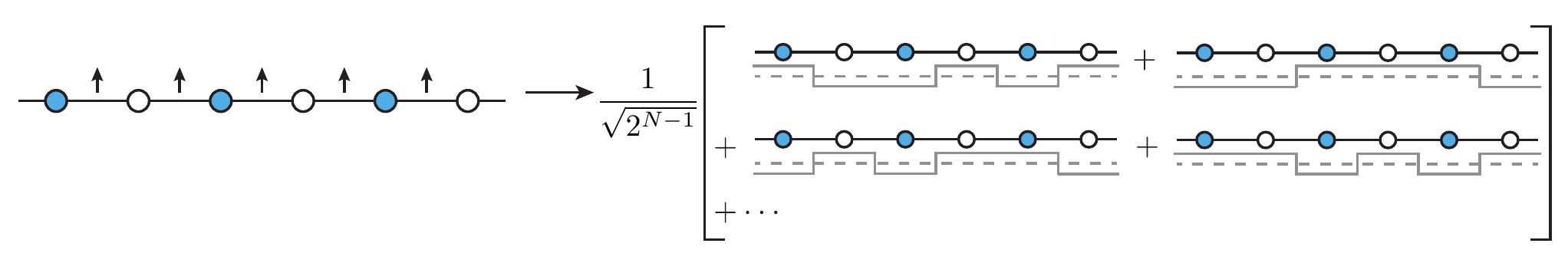}
	\caption{Schematic picture showing the transformation of the initial state into a dual representation. On the left is an initial state with fermions in a charge density wave (filled sites -- blue, empty sites -- white). The bond spins are polarized along the $z$-axis. The dual state (right panel) has the fermions in the same configuration, but the wave-function is an equal superposition of all charge configurations. In the dual representation, the charge sector leads to exact averaging over the binary potential, which is shown in grey.}\label{fig: state transformation}
\end{figure*}

\subsection{Duality Mapping}

In our previous work~\cite{Smith2017,Smith2017_2} we found a mapping that reveals an equivalence between charge configurations $\{q_i = \pm1\}$ and configurations of on-site potentials for the fermions. Here we explain this mapping in greater detail and for the more general class of models described by Eq.~\eqref{eq: H general}.

We proceed by a duality transformation of the operators $\sigma$, defining spin-1/2 operators $\tau$ which live on the sites of the lattice,
\begin{equation}\label{eq: duality}
\hat{\tau}^z_j = \hat{A}_j, \qquad \hat{\tau}^x_j\hat{\tau}^x_k = \hat{\sigma}^z_{jk},
\end{equation}
where indices $j$ and $k$ correspond to nearest neighbour sites, see Fig.~\ref{fig: 2D model}. We have to choose one of the disconnected subspaces of the model. These disconnected subspaces can be enumerated by another set of conserved quantities defined as products of $\hat{\sigma}^z$ along closed loops on the lattice. These conserved quantities can be expressed in terms of plaquette operators, $\hat{B}_p$, defined on the irreducible plaquettes of the lattice (see Fig.~\ref{fig: 2D results}) and Wilson loop operators $\hat{\Gamma}_n$,
\begin{equation}
\hat{B}_p = \prod_{\text{plaquette } p} \hat{\sigma}^z_{jk}, \qquad  \hat{\Gamma}_n = \prod_{\la jk\ra \in \gamma_n} \hat{\sigma}^z_{jk},
\end{equation}
where $\gamma_n$ is any closed path that winds around a torus and which cannot be written as a product of plaquette operators. For concreteness, we give two examples: in a 1D periodic chain there is only one such operator which is the loop around the entire system $\prod\hat{\sigma}^z$. This is equivalent to a statement that a number of domain walls modulo 2 is conserved. On a 2D torus we have $\hat{B}_p$ on all square plaquettes of the lattice and the two Wilson loops around the two periodic directions. Importantly, as well as commuting with the Hamiltonian, these operators commute with the generators of the $\mathbb{Z}_2$ gauge symmetry $\hat{q}_i$. The eigenvalues $\pm 1$ of these operators label subspaces which are disconnected under gauge transformations.

The duality mapping in Eq.~\eqref{eq: duality} forces the choice $\hat{B}_p = 1$ on all plaquettes and all $\hat{\Gamma}_n = 1$. For a discussion of different plaquette sectors, see Ref.~\cite{Prosko2017}. In all but one dimension with open boundary conditions we also have a global constraint which is due to the product of all star operators being equal to the identity, i.e. $\prod_{\text{all } i} \hat{A}_i = 1$. On the other side of the duality mapping this amounts to the constraints
\begin{equation}\label{eq: global constraints}
\prod_{\text{all } i} \hat{\tau}^z_i = 1, \qquad \prod_{\text{all } i}\hat{q}_i = (-1)^{\hat{N}_f},
\end{equation}
where $\hat{N}_f = \sum_i \hat{n}_i$ is the total fermion number.


In terms of the $\tau$ spins the Hamiltonian assumes the form
\begin{equation}
\hat{H} = - J \sum_{\la j k \ra} \hat{\tau}^x_j \hat{\tau}^x_k \hat{f}^\dagger_j \hat{f}_k - h \sum_j \hat{\tau}^z_j.
\end{equation}
Although this Hamiltonian is equivalent to Eq.~\eqref{eq: H general} only on a restricted Hilbert space, we will not use notation to distinguish between the two. This form is identical to the one exposed for the 1D chain~\cite{Smith2017,Smith2017_2}, but here the nearest neighbour connectivity can be described by any graph. In this form we can identify local conserved quantities $\hat{q}_j = \hat{\tau}^z_j (-1)^{\hat{n}_j}$ with $\hat{n}_j = \hat{f}^\dagger_j \hat{f}_j$. These \emph{charges} commute with the Hamiltonian and amongst themselves. The charges are precisely those that generate the gauge symmetry identified in the original degrees of freedom. 

Finally, by a change of variables $\hat{c}_j = \hat{\tau}^x_j \hat{f}_j$, the Hamiltonian can be written in terms of conserved charges and the spinless fermions $\hat{c}$:
\begin{equation}\label{eq: H qs}
\hat{H} = -J \sum_{\la j k \ra} \hat{c}^\dagger_j \hat{c}_k + 2h \sum_j \hat{q}_j (\hat{c}^\dagger_j \hat{c}_j - 1/2),
\end{equation}
where we have used the fact that $\hat{n}_j = \hat{f}^\dagger_j\hat{f}_j = \hat{c}^\dagger_j\hat{c}_j$, since $(\hat{\tau}^x_j)^2 = 1$. The canonical commutation relations $\{\hat{c}^\dagger_j,\hat{c}_k\} = \delta_{jk}$ can be similarly verified. For a given charge configurations -- that is in the subspace of fixed $\{q_j\} = \pm 1$ -- the Hamiltonian~\eqref{eq: H qs} describes a tight-binding model with a binary potential whose sign is set at each site by the value of $q_j$. Note that we recover the Falicov-Kimball model if we impose the global constraint $\sum_i (\hat{q}_i + 1) |\Psi\ra = \mu |\Psi\ra$, where $\mu$ is an integer that corresponds to the chemical potential for the static auxiliary fermions $\hat{g}_j$, defined via $\hat{q}_i = \hat{g}_i^\dagger\hat{g}_i$.

As well as understanding how the operators transform under the mapping, we must also make an identification between the eigenstates. Let us consider tensor product states of the form $|\Psi\ra_{\sigma,f} = |S\ra_\sigma \otimes |\psi \ra_f$, which we wish to identify with a state $|\Psi\ra_{\tau,c}$ in the Hilbert space of the $\tau$ and $c$ degrees of freedom, and in turn with $|\Psi\ra_{q,c}$. If for the fermion states we choose the Fock states, i.e., $|\psi\ra_f = \hat{f}^\dagger_j \cdots \hat{f}^\dagger_l |vacuum\ra$, then these states take the same form for the $c$ fermions, and we will drop the subscript in the following. Without loss of generality, let us consider spins in the z-polarized state $| \!\uparrow \uparrow \uparrow \cdots\ra_\sigma$ -- any other spin state in the sector defined by all $\hat{B}_j = 1$ can be reached via application of star operators $\hat{A}_j$. By the duality transformation and the global constraint of Eq.~\eqref{eq: global constraints} we have that
\begin{equation}
\hat{\tau}^x_{i}\hat{\tau}^x_j |\!\uparrow\uparrow\uparrow \cdots \ra_\sigma  = \prod_{\text{all } i} \hat{\tau}^z_i |\!\uparrow\uparrow\uparrow \cdots \ra_\sigma = |\!\uparrow\uparrow\uparrow \cdots \ra_\sigma,
\end{equation}
and we make the correspondence between states $|\!\uparrow\uparrow\uparrow \cdots \ra_\sigma = \frac{1}{\sqrt{2}} (|\!\rightarrow\rightarrow\cdots\ra_\tau + |\!\leftarrow\leftarrow\cdots\ra_\tau)$.
Therefore we can generally make identifications of the form $|S\ra_\sigma \otimes |\psi\ra \propto |S\ra_\tau \otimes |\psi \ra$, where $|S\ra_{\sigma(\tau)}$ has a definite local $z$($x$)-component of spin, and $|\psi\ra$ has a definite local occupation. 


Now we can express these tensor product states in terms of conserved charges in place of the $\tau$ spins. For a $z$-polarized state this proceeds as follows
\begin{equation}
|\!\uparrow\uparrow\uparrow \cdots \ra_\sigma \otimes |\psi\ra = \frac{1}{\sqrt{2^{N-1}}} \sideset{}{'}\sum_{\{\tau_i\} = \uparrow,\downarrow} |\tau_1,\tau_2 \cdots \ra_\tau \otimes |\psi\ra,
\end{equation}
where we have identified $|\!\rightarrow\ra_\tau = (|\!\uparrow \ra_\tau + |\!\downarrow\ra_\tau)/\sqrt{2}$, for each $\tau$ spin, and the prime indicates that the sum runs over all configurations satisfying the constraint~\eqref{eq: global constraints}. Let us consider a single state in this sum $|\tau_1 \tau_2 \cdots \ra_\tau \otimes |\psi\ra$, then since the fermion state is a simple tensor product of site occupation, this can be rewritten as
\begin{equation}
|\tau_1(-1)^{n_1}, \tau_2(-1)^{n_2}, \cdots \ra_{q} \otimes |\psi \ra.
\end{equation}
The occupation numbers for the fermion state are fixed and thus only contribute a common sign structure to the charge configuration. Since we sum over all $\tau$ configurations, all with a positive weight, this equates to a sum over all charge configurations
\begin{equation}\label{eq: q state}
|\uparrow \uparrow \cdots \ra_\sigma \otimes |\psi\ra = \frac{1}{\sqrt{2^{N-1}}}\sideset{}{'}\sum_{\{q_j\}=\pm 1} |q_1,q_2, \cdots, q_N\ra \otimes |\psi\ra,
\end{equation}
where again the primed sum indicates the constraint~\eqref{eq: global constraints}.
This transformation of states is shown schematically in Fig.~\ref{fig: state transformation}. The fact that all of the weights are equal and positive is important for this final form, otherwise there would be a sign structure that depends both on the spin and the fermion configuration. Other spin states in the same spin sector can be accessed through the application of star operators.

\subsection{Emergent disorder and disorder averaging}

In the previous section we showed a transformation to the Hamiltonian~\eqref{eq: H qs} which has an effective binary potential, and that the states in the dual configuration are superpositions of the states with a given charge configuration, with the sign of the charges generating a potential for the fermions. 
Throughout this paper -- except briefly in Section~\ref{sec: toric code} -- we will consider the quenched initial states $|\Psi\ra = |\!\uparrow\uparrow\uparrow\cdots \ra_\sigma\otimes |\psi\ra$ with $z$-polarized spins and a selection of fermion Slater determinants. In order to make a connection with the localization problem, the final step is to show that expectation values of observables with respect to these initial states amount to averages over effective disorder.

Let us for concreteness consider a spin expectation value 
\begin{equation}\label{eq: q correlator}
\begin{aligned}
&\!\!\la \Psi | \hat{\sigma}^z_{jk}(t)  |\Psi\ra\\ &= \frac{1}{2^{N-1}}\! \sideset{}{'}\sum_{\{s_l\},\{q_m\}=\pm1} \!\!\!\la \psi | \la s_1, \cdots | e^{i\hat{H}t} \hat{\tau}^x_j \hat{\tau}^x_k e^{-i\hat{H}t} |q_1, \cdots \ra  | \psi \ra.
\end{aligned}
\end{equation}
In order to simplify expressions we introduce the fermion Hamiltonian 
\begin{equation}\label{eq: H q}
\hat{H}({\{q_j\}}) = -J \sum_{\la j k \ra} \hat{c}^\dagger_j \hat{c}_k + 2h \sum_j q_j (\hat{c}^\dagger_j \hat{c}_j - 1/2).
\end{equation}
The difference with Eq.~\eqref{eq: H qs} is that this Hamiltonian \ref{eq: H q} acts only in the fermion subspace and the $q_j$ are no longer operators -- the configuration $\{q_j\}=\pm 1$ is specified. Equation~\eqref{eq: q correlator} can then be written as
\begin{equation}\label{eq: spin average}
\begin{aligned}
\!\la \Psi |& \hat{\sigma}^z_{jk}(t)  |\Psi\ra\\ 
&= \frac{1}{2^{N-1}} \sideset{}{'}\sum_{\{q_i\} = \pm 1} \la \psi | e^{i\hat{H}(\{q_i\})t}
e^{-i\hat{H}(\{q_i:\bar{q}_j,\bar{q}_k\}) t} | \psi \ra,
\end{aligned}
\end{equation}
where $\bar{q}_j$ signifies that the value of charge $q_j$ in the Hamiltonian has been reversed. This reversal of the charge arises from commuting $\tau$ operators past the time evolution operator. Note that one can also then remove the second sum over charges because of charge conservation. Similar arguments can be used to show that all correlators of this form reduce to fermion correlators averaged over all charge configurations, which amounts to all disorder configurations in the Hamiltonian~\eqref{eq: H q}~\cite{Paredes2005,Enss2016}. The correlators can be efficiently computed using determinants, see Appendix~\ref{ap: determinant}. It is important to note that, as in Eq.~\eqref{eq: spin average}, the expressions for the correlators that we obtain are distinct from, e.g., the fermion correlators of a tight-binding model with disorder. For instance, the Green's function 
\begin{equation}
\begin{aligned}
\!\!\!\!\!\la \hat{f}^\dagger_j&(t) \hat{f}_k(0) \ra \\
&= \frac{1}{2^{N-1}} \sideset{}{'}\sum_{\{q_i\} = \pm 1} \la \psi | e^{i\hat{H}(\{q_i\})t} \hat{c}^\dag_j
e^{-i\hat{H}(\{q_i:\bar{q}_j\}) t} \hat{c}_k | \psi \ra,
\end{aligned}
\end{equation}
does not correspond to averaging over disorder configurations for the Green's functions $\la \hat{c}^\dagger_j(t) \hat{c}_k(0) \ra$ because of the flipped charges between the forward and backward time evolution. In this respect the correlators that we obtain are similar to the ones appearing in the X-ray edge problem and the dynamical structure factor for the honeycomb Kitaev model~\cite{BaskaranExact,PRL}. On the other hand, density averages and correlators do equate to simple disorder-averaging without flipped charges.

\subsection{Defect attachment in the toric code model}\label{sec: toric code}

Before moving on to study the physics of localization in our model, let us first consider the Hamiltonian~\eqref{eq: H general} defined on a 2D square lattice with periodic boundary conditions. This becomes equivalent to Kitaev's toric code~\cite{KitaevTQM} -- with plaquette dynamics frozen -- coupled to spinless fermions. Note that one way of introducing dynamical defects into the toric code is by adding transverse field terms $\sum_{\la jk \ra} \hat{\sigma}^z_{jk}$. Here we briefly outline the dynamics induced by our coupling to fermions.

For our discussion of localization, we study the initial states of spins polarized along the $z$-axis. However, let us for a moment focus on spin states describing the ground state of the toric code, that is
\begin{equation}
\hat{A}_j |S_0 \ra = |S_0\ra, \qquad \hat{B}_p |S_0 \ra = |S_0 \ra.
\end{equation}
We note that our choice of duality transformation is consistent with this ground state, and it fixes Wilson loop operators to be $\hat{\Gamma}_1 = \hat{\Gamma}_2 = 1$, thus it uniquely chooses one of the four degenerate ground states (for $J=0$) of the toric code. To access other ground state sectors we can modify the duality transformation~\eqref{eq: duality} by defining a vertical and horizontal line (going through bonds) over which the implicit definition of $\hat{\tau}^x$, picks up a sign $\Gamma_1, \Gamma_2 = \pm 1$, respectively. More explicitly, one can define
\begin{equation}
\hat{\tau}^x_j \hat{\tau}^x_k = (\Gamma_1)^{\delta^1_{jk}}(\Gamma_2)^{\delta^2_{jk}} \hat{\sigma}^z_{jk},
\end{equation}
where $\delta^{1(2)}_{jk}$ is $1$ when the bond $\la j k \ra$ crosses a vertical(horizontal) reference line, and $0$ otherwise. Note that this choice changes the action of the Wilson loop operators but not $\hat{B}_p$, since any plaquette crosses any line an even number of times.

Having chosen the initial spin configuration, we can consider the coupling to fermions.
For a simple tensor product state $|S_0\ra\otimes |\psi\ra$, this maps to
\begin{equation}
|\uparrow \uparrow \cdots \ra_\tau \otimes |\psi\ra = |(-1)^{\hat{n}_1}, (-1)^{\hat{n}_2},\cdots\ra_q \otimes |\psi\ra,
\end{equation}
that is, for an initial fermion state of definite local occupation, the charge configuration is uniquely specified by the parities of fermion occupation numbers on each site. The Hamiltonian then takes a simple form
\begin{equation}
\hat{H}_\text{TC} = -J \sum_{\la j k \ra} \hat{c}^\dagger_j \hat{c}_k + 2h \sum_j q_j (\hat{c}^\dagger_j \hat{c}_j - 1/2),
\end{equation}
where in contrast to Eq.~\eqref{eq: H qs}, the potential given by $q_j$ is fixed and equal to $-1$ if there is a fermion on this site in the initial state, and $+1$ if the site is empty. If we consider the limit $h\gg J$, then the fermions lie at the bottom of large potential wells, and fermion hopping is suppressed, and we recover the static toric code. The form of the conserved charges is essentially a statement that defects in the toric code are attached to fermions (or holes).

Let us now consider excitations of this model in the limit of $h\gg J$. Addition of a star defect on a site amounts to flipping a $\tau$-spin and corresponds to changing the sign of the potential on the same site. This defect is then free to move in a restricted geometry on the lattice that is determined by the sites occupied by fermions, since the defect is attached to a fermion (hole). This geometry corresponds to a connected region of the lattice which has the opposite fermion parity to the fermion/hole attached to the defect. This can be understood as site percolation problem for the defects, and we will encounter it again in Section~\ref{sec: delocalization}. Importantly on a square lattice the percolation threshold is $p_c \approx 0.5927$, which means that for fermions at half filling and in a random configuration, the defects are localized.

\begin{figure*}[!tb]
	\centering
	\subfigimg[width=.335\textwidth,valign=b]{\hspace*{0pt} \textbf{(a)}}{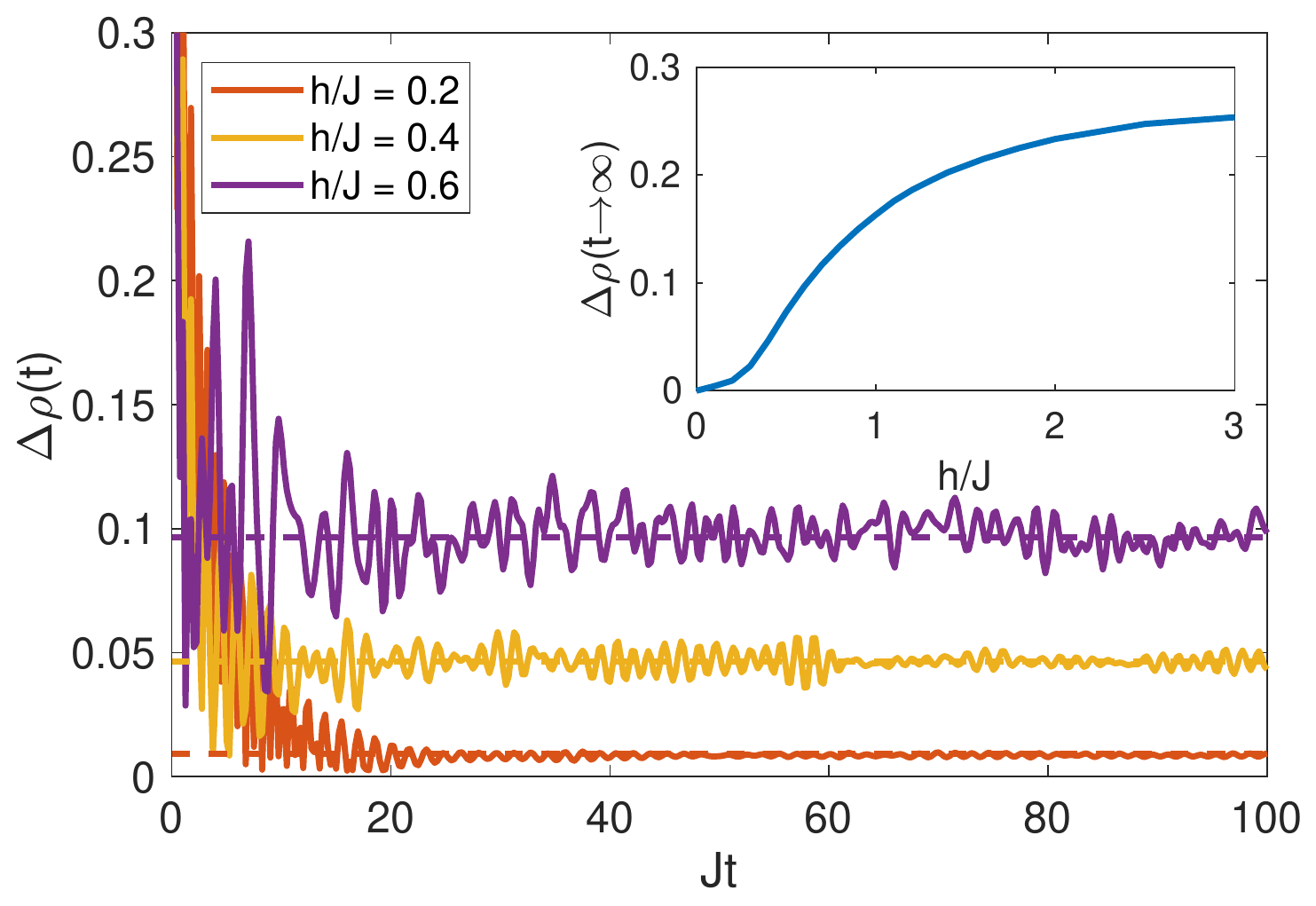}
	\subfigimg[width=.147\textwidth,valign=b]{\hspace*{0pt} \textbf{(b)}}{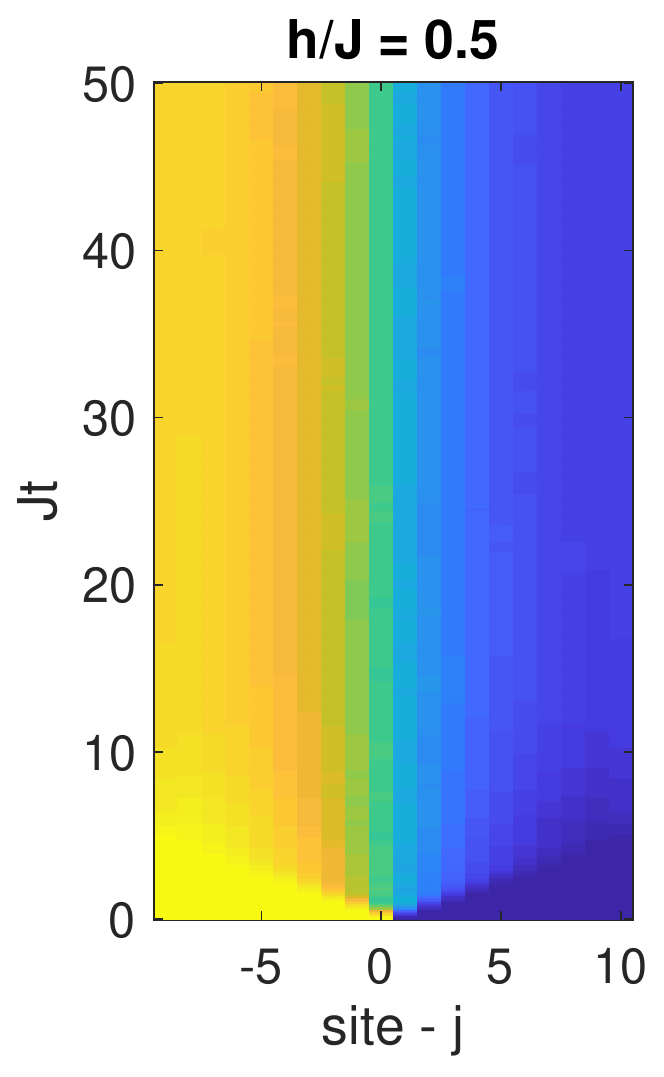}
	\subfigimg[width=.147\textwidth,valign=b]{\hspace*{0pt} \textbf{(c)}}{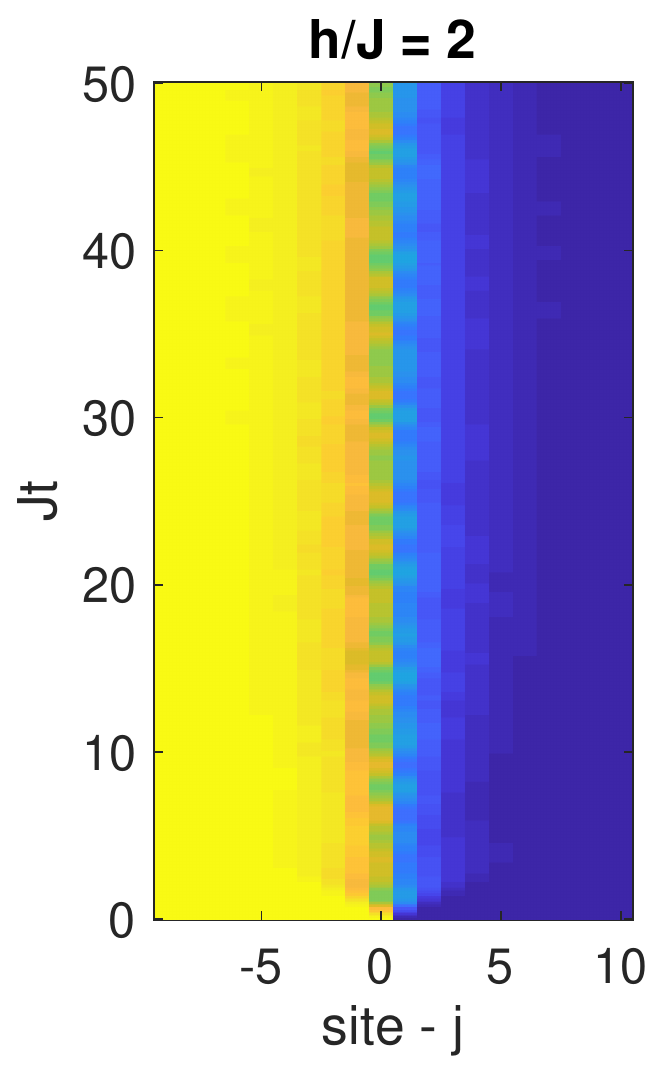}
	\subfigimg[width=.335\textwidth,valign=b]{\hspace*{0pt} \textbf{(d)}}{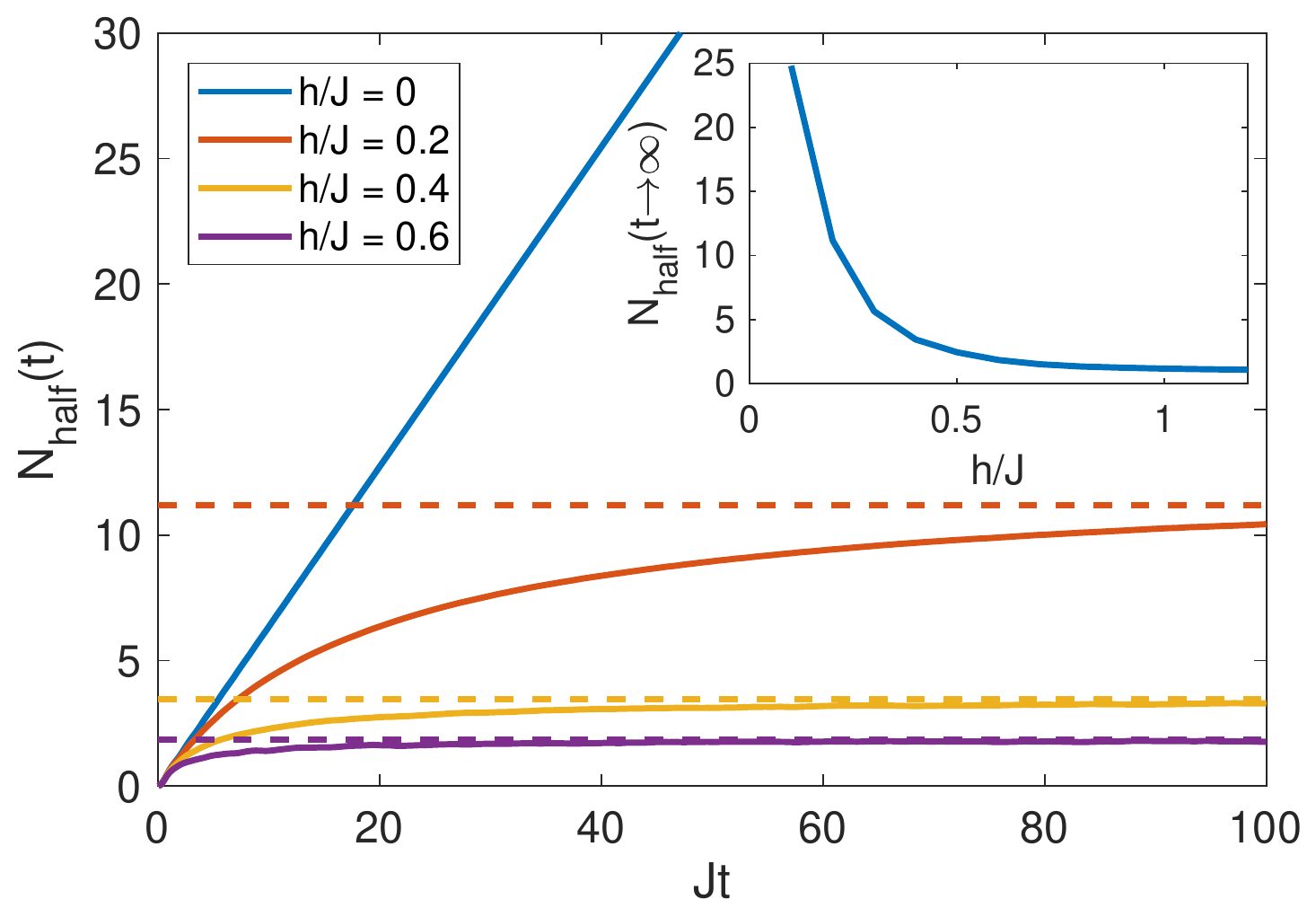}
	\caption{Results for the time-evolution of the fermion subsystem in 1D. (a) Persistence of a charge density wave measured by the density imbalance between neighbouring sites, $\Delta\rho$, see Eq.~\eqref{eq: delta rho}, for $N=200$ sites with periodic boundary conditions. The inset shows results in the long-time limit, $Jt = 10^9$ as a function of effective disorder strength $h/J$. (b) and (c) Spreading of a domain wall (in the initial configuration) for $h/J = 0.5$ and $h/J = 2$ respectively for the system with $N=20$ sites and open boundary conditions. Local density is shown (filled--yellow, empty--blue). (d) Time evolution of particle occupation numbers on the right-half of the system $N_\text{half}$ from an initial domain wall configuration for $N=200$ sites. Inset shows results for the long time limit as a function of $h/J$. All calculations are performed using the determinant method of Appendix~\ref{ap: determinant}.}\label{fig: 1D results}
\end{figure*}

Let us now return to the discussion of the plaquette operators. We can include the latter in our model and through the symmetries of the model do exactly the same procedure of attaching defects to fermions. If we consider a dual lattice which is a square lattice with sites at the centres of plaquettes, then the plaquette operators become star operators and vice versa. We can then add a second fermion species on this dual lattice which will be attached to plaquette defects. If we denote fermions attached to star defects $\hat{a}$ and those attached to plaquette defects $\hat{b}$, then the toric code model with fermion attachment reads
\begin{equation}
\begin{aligned}
\hat{H}_\text{TC} = &-h_A \sum_s \hat{A}_s -h_B \sum_p \hat{B}_p\\ 
&- J_A \sum_{\la i j \ra_s} \hat{\sigma}^z_{j,k} \hat{a}^\dag_i \hat{a}_j - J_B \sum_{\la i j \ra_p} \hat{\sigma}^x_{j,k} \hat{b}^\dag_i \hat{b}_j,
\end{aligned}
\end{equation}
where $\la i j \ra_s$ denotes nearest neighbours on the original lattice, and $\la i j\ra_p$ denotes those on the dual lattice. The conserved charges are $\hat{q}^A_j = \hat{A}_j (-1)^{\hat{a}^\dag_j\hat{a}_j}$ and $\hat{q}^B_j = \hat{B}_j (-1)^{\hat{b}^\dag_j\hat{b}_j}$, respectively. We now have a full toric code with defects attached to fermions. We are then able to control the dynamical behaviour of the defects by choosing the initial configuration and occupation numbers for the fermions.

\section{Localization}\label{sec: localization}

In this section we discuss the localization behaviour of the model defined by Eq.~\eqref{eq: H general} in the case of a 1D chain and a 2D square lattice. Since the model can be mapped to free fermions, we can calculate correlators using determinants as explained in Appendix~\ref{ap: determinant}. This approach allows us to study systems with $\sim10^2-10^3$ sites. For entanglement entropy calculations we require the full density matrix and resort to exact diagonalization in 1D for up to $N=12$ sites. To calculate the density of states, we use the kernel polynomial method~\cite{Weisse2006}, see Appendix~\ref{ap: KPM}, which can be used for systems of order $10^5$--$10^6$ sites. Localization lengths are computed using a standard transfer matrix approach~\cite{Kramer1993} described in Appendix~\ref{ap: transfer matrix}.

\subsection{Dynamical localization in 1D}

The quench problem that we study here has initial states with bond spins polarized along the $z$-axis $|\!\uparrow\uparrow\uparrow\cdots \ra$, and with fermions in one of the following Slater determinant states:\\
(i) \textit{Charge density wave} described by fermions in a Fock state with occupation numbers $|\cdots 1010\cdots\ra$. We will probe the memory of this initial state via the nearest-neighbour density imbalance
\begin{equation}\label{eq: delta rho}
\Delta\rho(t) = \frac{1}{\widetilde{N}}\sum_j |\la \Psi | \hat{n}_j(t) - \hat{n}_{j+1}(t) | \Psi \ra|, 
\end{equation}
where $\widetilde{N}=N-1,N$, for open and periodic boundary conditions respectively. This measure was used e.g.~to identify the MBL transition in cold atom experiments, see Ref.~\cite{Schreiber2015}; \\
(ii) \textit{Domain wall} configuration with the left half of the chain filled and the right half empty $|\cdots111000\cdots \ra$. In order to quantify localization in this case we  measure the total number of particles in the right half of the system (which is empty in the initial state), which tells us how many particles make it across the domain wall,
\begin{equation}
N_\text{half}(t) = \sum_{j \in \text{ right half}} \la \Psi| \hat{n}_j(t) |\Psi \ra.
\end{equation}
This observable, as well as the long-time fermion density distribution, reveal the extent to which the fermions are localized. A similar measurement was used to identify the MBL transition  in 2D, see the cold atom experiments of Ref.~\cite{Choi2016}, and in theoretical work as a dynamical measure of localization, see Ref.~\cite{Hauschild2016}.

\begin{figure}[bh!]
	\centering
	\subfigimg[width=.46\textwidth]{\hspace*{0pt} \textbf{(a)}}{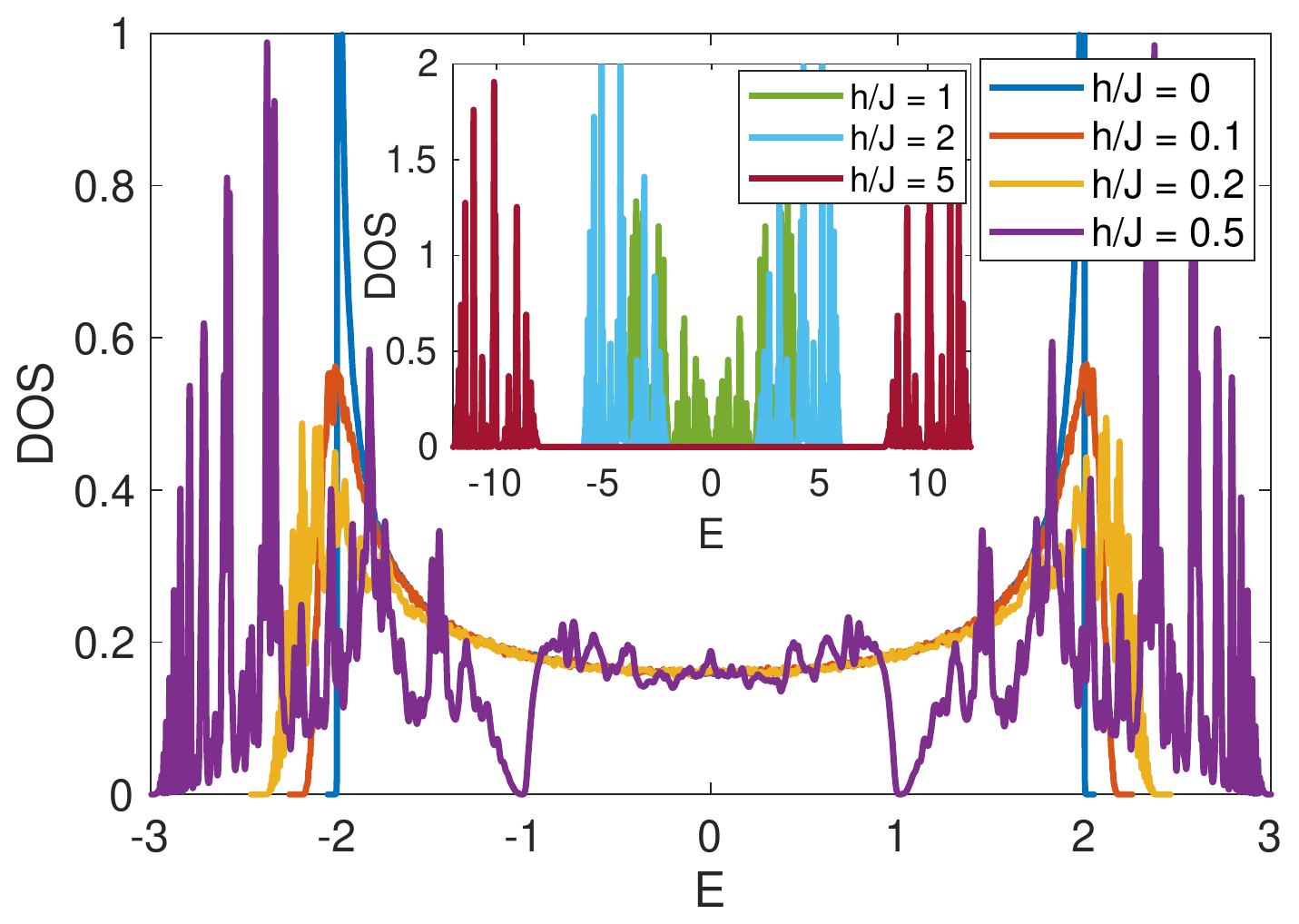}
	\subfigimg[width=.45\textwidth]{\hspace*{0pt} \textbf{(b)}}{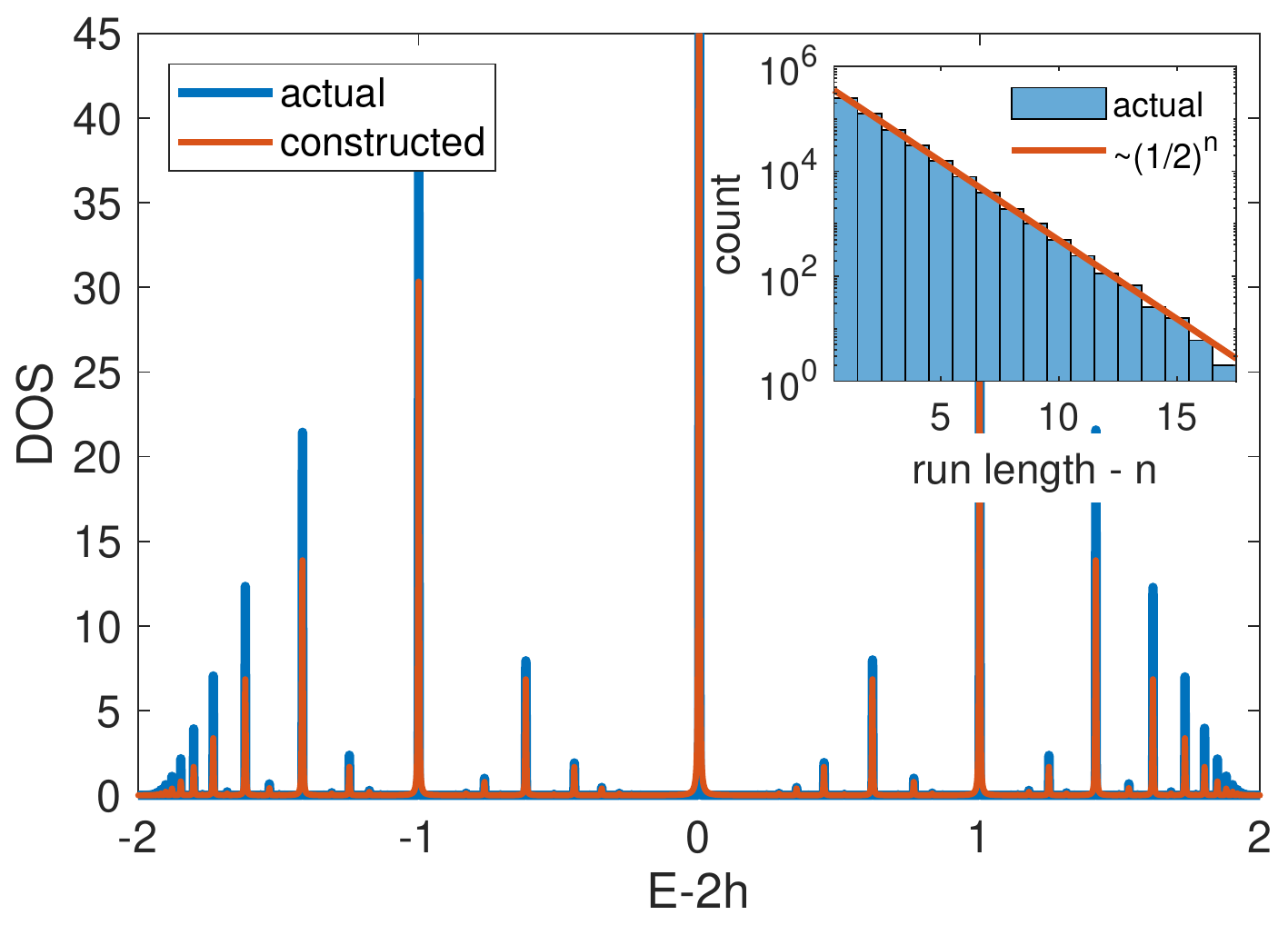}
	\caption{Density of states for the 1D chain~\eqref{eq: H q}. (a) DOS for different values of $h/J$. Inset shows the DOS for values of $h>J$ (where there is a gap in the DOS). (b) DOS for a very large value of $h/J = 500$. The energy is offset by $2h$ and we focus on one of the two sub-bands that form for large $h$. The DOS is computed using the kernel polynomial method (see Appendix~\ref{ap: KPM}), shown in blue. We compare this with the DOS constructed using Eq.~\eqref{eq: constructed DOS}, shown in red. Inset in panel (b) shows a comparison of the observed distribution of chain lengths with the corresponding distribution in the thermodynamic limit.}\label{fig: 1D DOS}
\end{figure}

Let us first consider the charge density wave initial state, where we measure the density imbalance $\Delta\rho(t)$, see Fig.~\ref{fig: 1D results}(a). While the latter decreases as a function of time from its initial value of $1$, at long times it approaches a non-zero value for all $h\neq 0$. Furthermore, the asymptotic value $\Delta\rho(t\rightarrow \infty)$ increases monotonically with $h/J$, as shown in the inset. This shows the memory of the initial state which is preserved in breaking of ergodicity due to emergent disorder. We also note that fluctuation in the asymptotic value of the density imbalance increase both in amplitude and longevity with increasing $h$. These fluctuations can be linked to the behaviour of the density of states shown in Fig.~\ref{fig: 1D DOS}(a), where we see that the DOS shows multiple spikes at the band edges that lead to resonances.

Next, let us consider the domain wall configuration shown in Fig.~\ref{fig: 1D results}(b-d). For $h/J < 1$ we observe initial linear spreading of the domain wall at short times. At longer times this linear spreading halts, and the density quickly approaches its limiting value with exponential tails set by the single particle localization length~\cite{Smith2017}, see Fig.~\ref{fig: 2D DOS}(b). For $h/J > 1$, we see a similar phenomenology except the halting is much more abrupt and the domain wall spreads over only a few sites, see Fig.~\ref{fig: 1D results}(c). The persistence of the domain wall can be most clearly quantified by $N_\text{half}(t)$: the number of particles that make it into the initially empty half of the system, as shown in Fig.~\ref{fig: 1D results}(d). We see that for $h\neq 0$ its behaviour deviates from linear growth which we observe for $h = 0$. The number of particles that make it across the domain wall is bounded showing that there is a remaining imbalance between the two halves of the system, and thus the memory of the initial domain wall remains at arbitrary long times.

\begin{figure}[tb!]
	\centering
	\includegraphics[width=.45\textwidth]{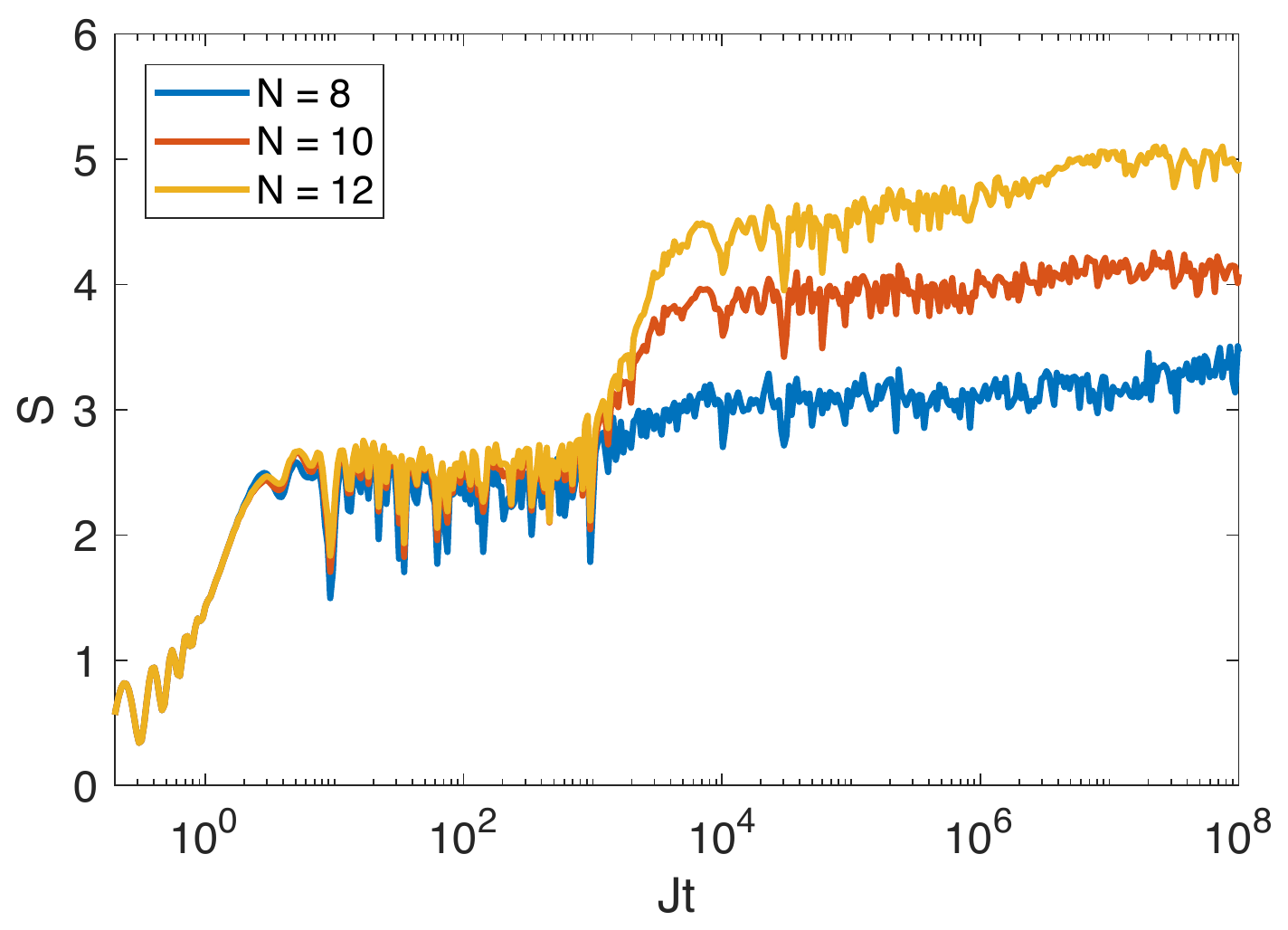}
	\caption{Time evolution of the bipartite von Neumann entanglement entropy in the 1D case. We start from a charge density wave initial state in a system with open boundary conditions and $h/J=10$. We partition the system into two halves along the central bond and results are computed using exact diagonalization.}\label{fig: Entanglement Pure}
\end{figure}

\begin{figure*}[!tb]
	\centering
	\subfigimg[width=.335\textwidth,valign=b]{\hspace*{0pt} \textbf{(a)}}{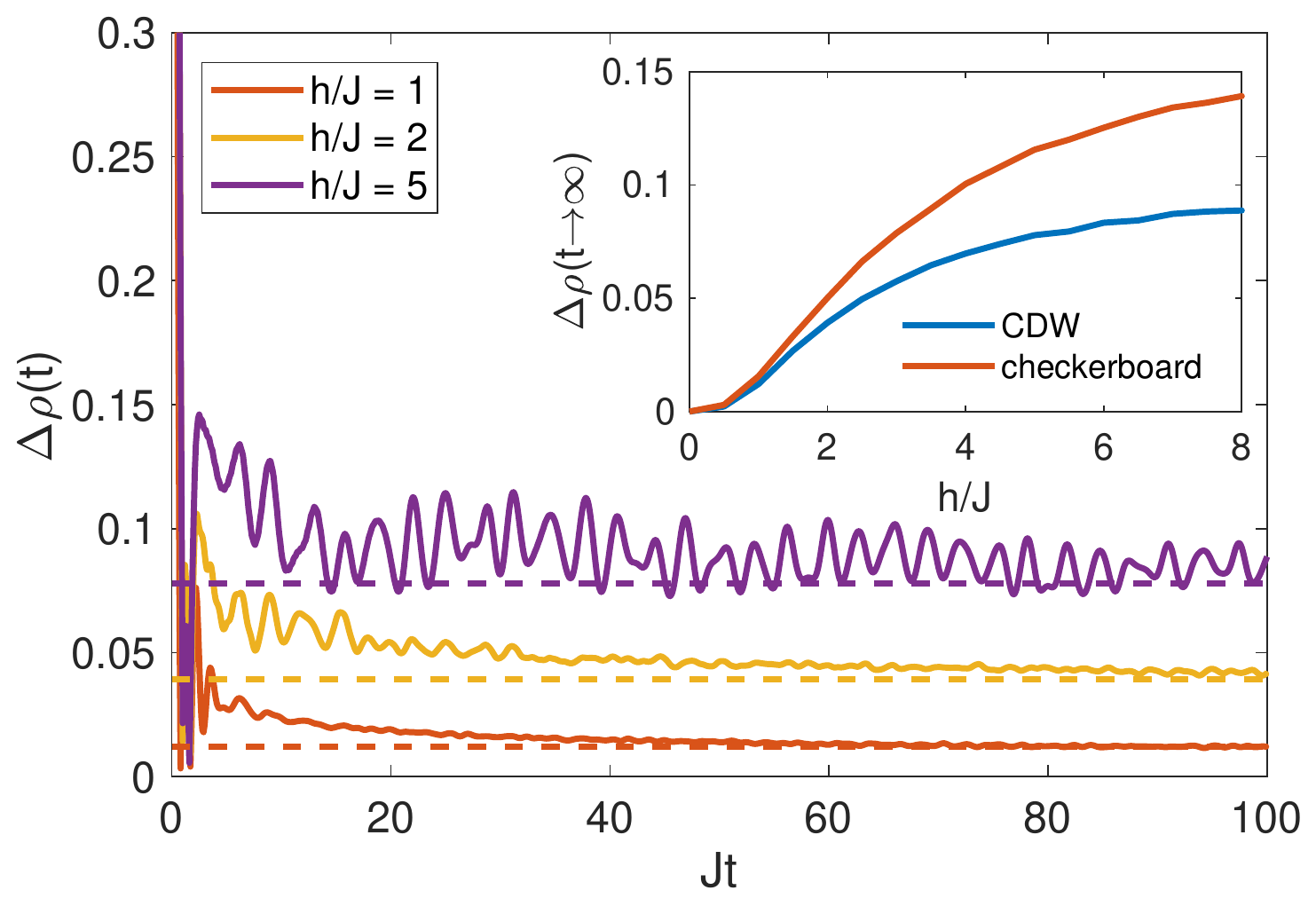}
	\subfigimg[width=.147\textwidth,valign=b]{\hspace*{0pt} \textbf{(b)}}{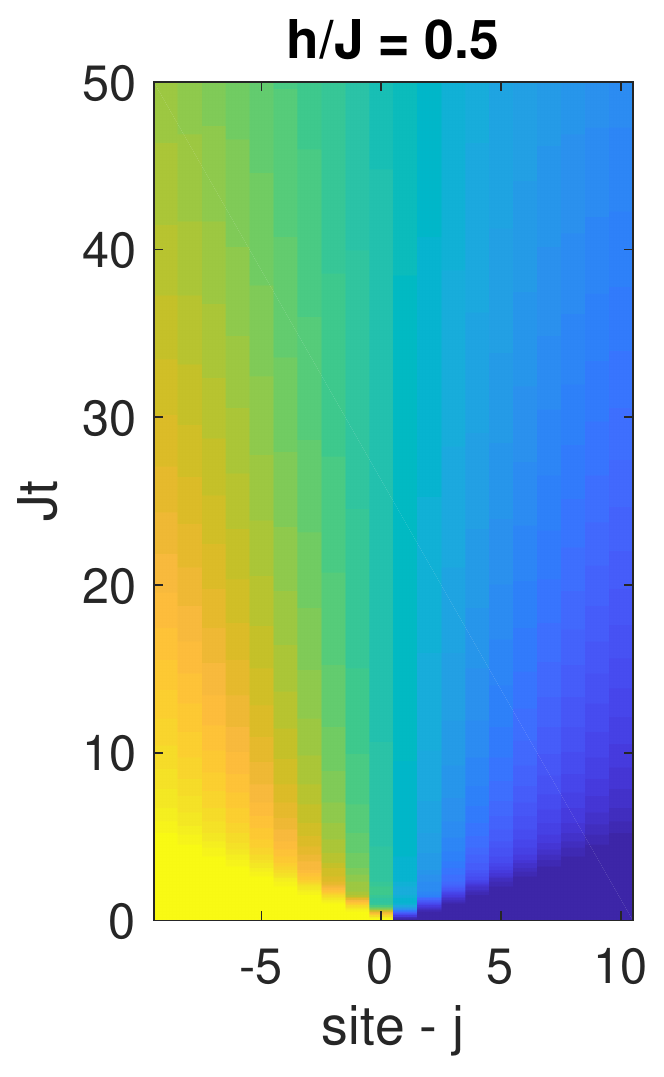}
	\subfigimg[width=.147\textwidth,valign=b]{\hspace*{0pt} \textbf{(c)}}{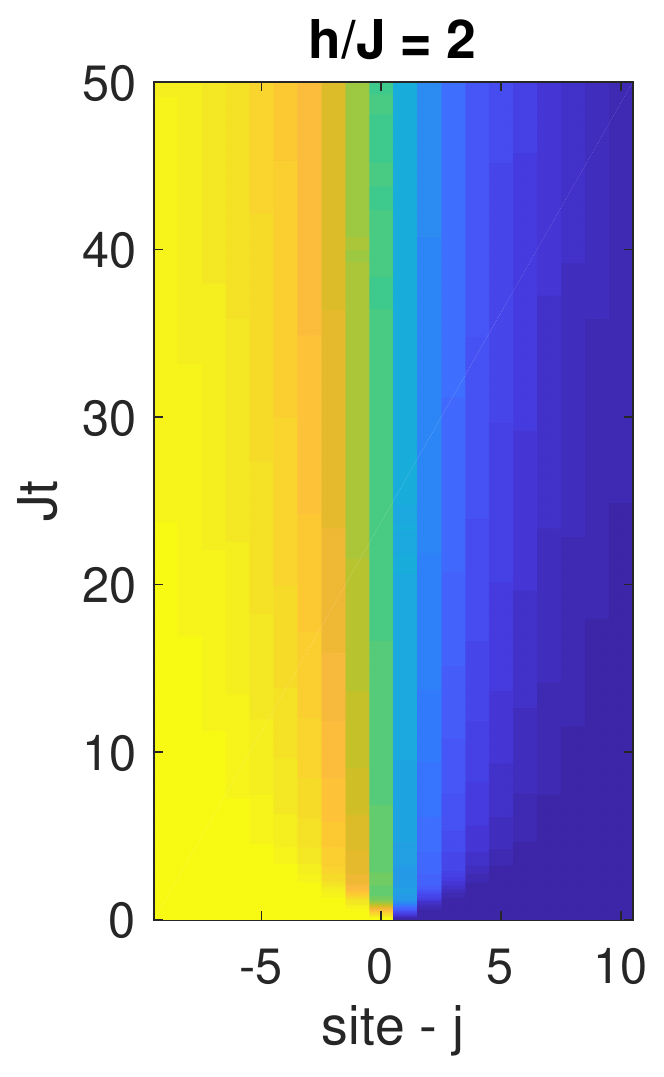}
	\;\;\subfigimg[width=.33\textwidth,valign=b]{\hspace*{0pt} \textbf{(d)}}{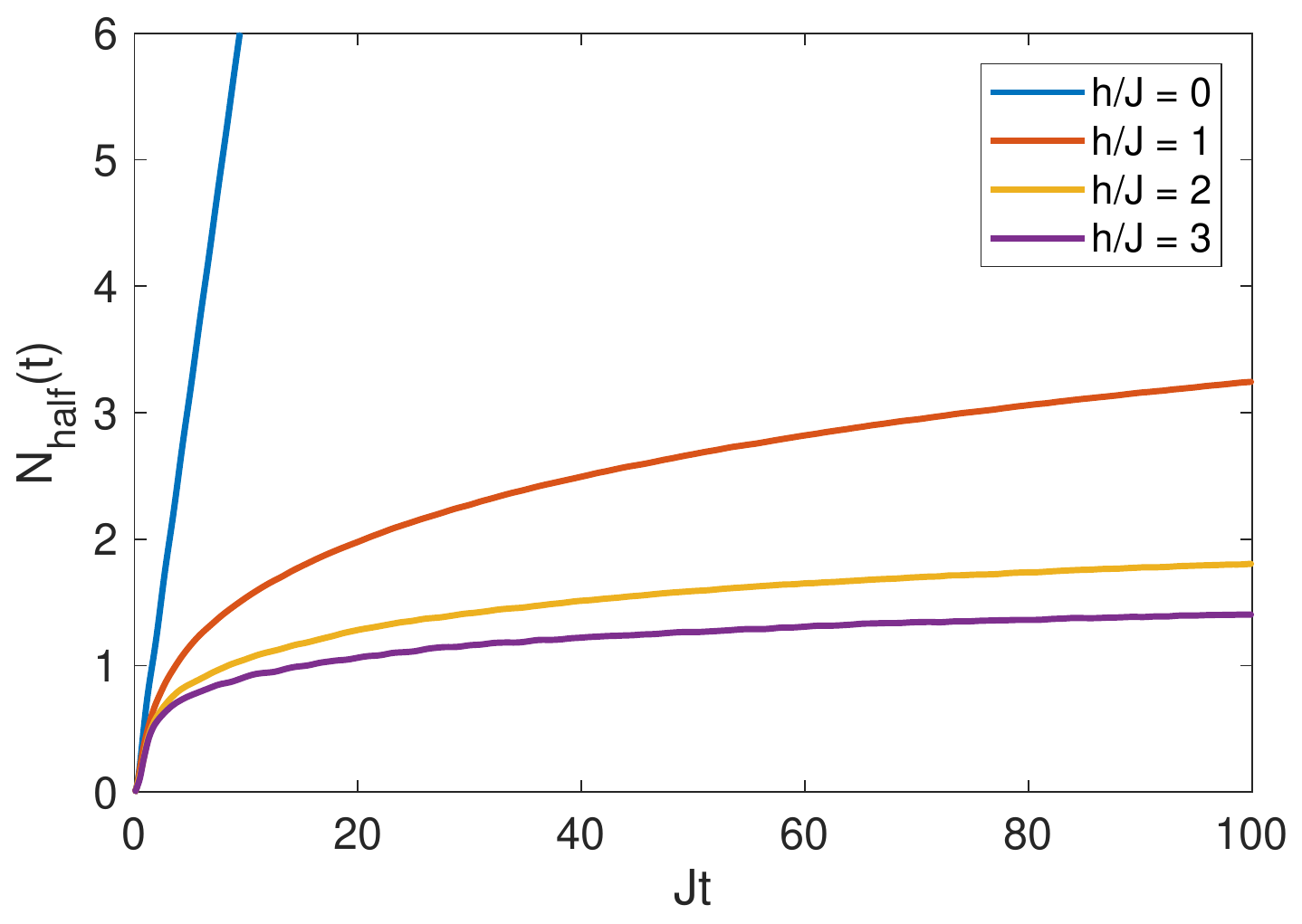}
	\caption{Time evolution of the fermion subsystem in 2D. (a) Density imbalance $\Delta\rho$ measured along a slice through the centre of the system with the initial state described by a charge density wave, see text. Inset shows the long-time limit for the charge density wave, and checkerboard initial states. (b) \& (c) Spreading of the domain wall for $h/J = 0.5$ and $h/J = 2$, respectively, measured along the slice through the centre of the system. (d) Number of particles $N_\text{half}$ along the centre in the initially empty half of the system. In (a-c) we use a square lattice with $N = 32\times32$ sites and in (d) $N= 30\times50$. Results are computed using the determinant method of Appendix~\ref{ap: determinant}.}\label{fig: 2D results}
\end{figure*}

\subsection{Strong disorder limit}\label{sec: strong disorder}

While we find localization for all values of $h$, we also observe a qualitative change in behaviour for $h/J>1$, which can be traced back to the binary nature of the effective disorder. See, for example, Refs.~\cite{Alvermann2005,Janarek2017} for further discussion of the differences between binary and continuous disorder. In our model, this difference is most evident in the bipartite von Neumann entanglement entropy, shown in Fig.~\ref{fig: Entanglement Pure}, where we observe a plateau with area law scaling. For both the CDW and the domain wall initial states, we also find that dynamical observables show larger amplitude and longer lived oscillations  with increasing $h/J$. This can be seen as fluctuations in $\Delta\rho$ in Fig.~\ref{fig: 1D results}(a) and side-to-side fluctuations near the domain wall in Fig.~\ref{fig: 1D results}(c).

To understand this behaviour let us consider the strong disorder limit, $h/J \gg 1$. Here one can  think of the 1D chain as a collection of finite-length disconnected chains, with the distribution of lengths, $l$, given by $~(1/2)^l$, as shown in the inset of Fig.~\ref{fig: 1D DOS}(b). In each isolated chain of length $l$, we then have $l$ single particle wave-functions and energy levels $E$. In this limit we can separate the Hamiltonian into $\hat{H}_h$ and $\hat{H}_J$ given by
\begin{equation}\label{eq: strong h Hamiltonian}
\begin{aligned}
\hat{H}_h &= 2h\sum_i q_i (\hat{c}^\dagger_i \hat{c}_i-1/2) - J\sum_{\la i j \ra : q_i = q_j} (\hat{c}^\dagger_i \hat{c}_j + \text{H.c}), \\
\hat{H}_J &= J\sum_{\la i j \ra : q_i = -q_j} (\hat{c}^\dagger_i \hat{c}_j + \text{H.c}).
\end{aligned}
\end{equation}
Here the sums are over nearest-neighbours satisfying the condition on the relative sign of $q$. Note that we omit an overall energy shift $h\sum_i q_i$ in the Hamiltonian, which is defined by a charge configuration. This shift does not affect the results since there is no matrix element between different charge sectors. The Hamiltonian $\hat{H}_h$ describes disconnected uniform tight-binding chains, and $\hat{H}_J$ corresponds to a hopping between these chains.

The DOS of the Hamiltonian $\hat{H}_h$ can be constructed using an ensemble of the energy levels for disconnected chains weighted by their probability distribution using the following equation
\begin{equation}\label{eq: constructed DOS}
g(\omega) \propto - \frac{1}{\pi} \Im{m} \lim_{\delta \rightarrow 0}\sum_{l=1}^\infty \sum_{E_l} \frac{(1/2)^l}{\omega - E_l + i\delta},
\end{equation}
where $E_l$ denote single-particle eigenvalues of the tight-binding Hamiltonian for a uniform chain of length $l$. In order to obtain the DOS numerically, we introduce a cutoff on the sum over $l$ and choose a finite broadening $\delta = 0.0015$. This form of the Hamiltonian and the corresponding DOS reveals two main features. First, for $h/J > 1$, we have a gap in the spectrum which splits into two sub-bands of bandwidth $4J$, centred at $\pm 2h$, see inset of Fig.~\ref{fig: 1D DOS}(a). Note that in the Falicov-Kimball model this corresponds to the Mott phase~\cite{Antipov2016}. Second, the DOS is characterised by a set of discrete peaks. In Fig.~\ref{fig: 1D DOS}(b), we compare the exact DOS centred around one of these sub-bands at $E = 2h$ for a large but finite system with large $h\gg J$, and the DOS constructed from Eq.~\eqref{eq: constructed DOS}, which shows good agreement. 

The fact that the DOS splits up into a discrete set, of which only a few carry the majority of the spectral weight, explains the observed fluctuations in our localization diagnostics together with the area-law plateau in the entanglement entropy. These features can be attributed to resonant processes between these few discrete states. Figure~\ref{fig: 1D DOS} shows that a similar structure persists, to some extent, below $h/J=1$. The effect of $\hat{H}_J$ on the DOS appears at second order in perturbation theory. This gives rise to the broadening of the spectrum, and provides a time-scale $\sim(h/J)^2$ which sets the lifetime of the area law plateau and of the fluctuations in the fermion density. The long-time area law appears because the spin subsystem relaxation time is given by this time-scale $\sim(h/J)^2$ arising from  resonant processes. On this time-scale the fermion subsystem explores the global structure of the charge distribution leading to a saturation of the entanglement between the charges and the fermions. Differences between binary and continuous disorder are discussed in e.g.~Ref.~\cite{Janarek2017}.

\subsection{Dynamical localization in 2D}

In two dimensions we will consider the Hamiltonian~\eqref{eq: H general} on a square lattice, see Fig.~\ref{fig: 2D model}. As in 1D we consider initial states with spins polarized along the $z$-axis. We study the initial states of fermions in one of the three following configurations:\\
(i) \textit{Charge density wave} with alternating occupation along one of the directions of the lattices and uniform occupation along the other (stripes);\\
(ii) \textit{Checkerboard pattern} with alternating occupation along both directions of the lattice;\\
(iii) \textit{Domain wall} configuration with one half of the system filled, and the other empty, such as studied in cold-atom experiments, see Ref.~\cite{Choi2016}.\\

\begin{figure}[bth!]
	\centering
	\subfigimg[width=.46\textwidth]{\hspace*{0pt} \textbf{(a)}}{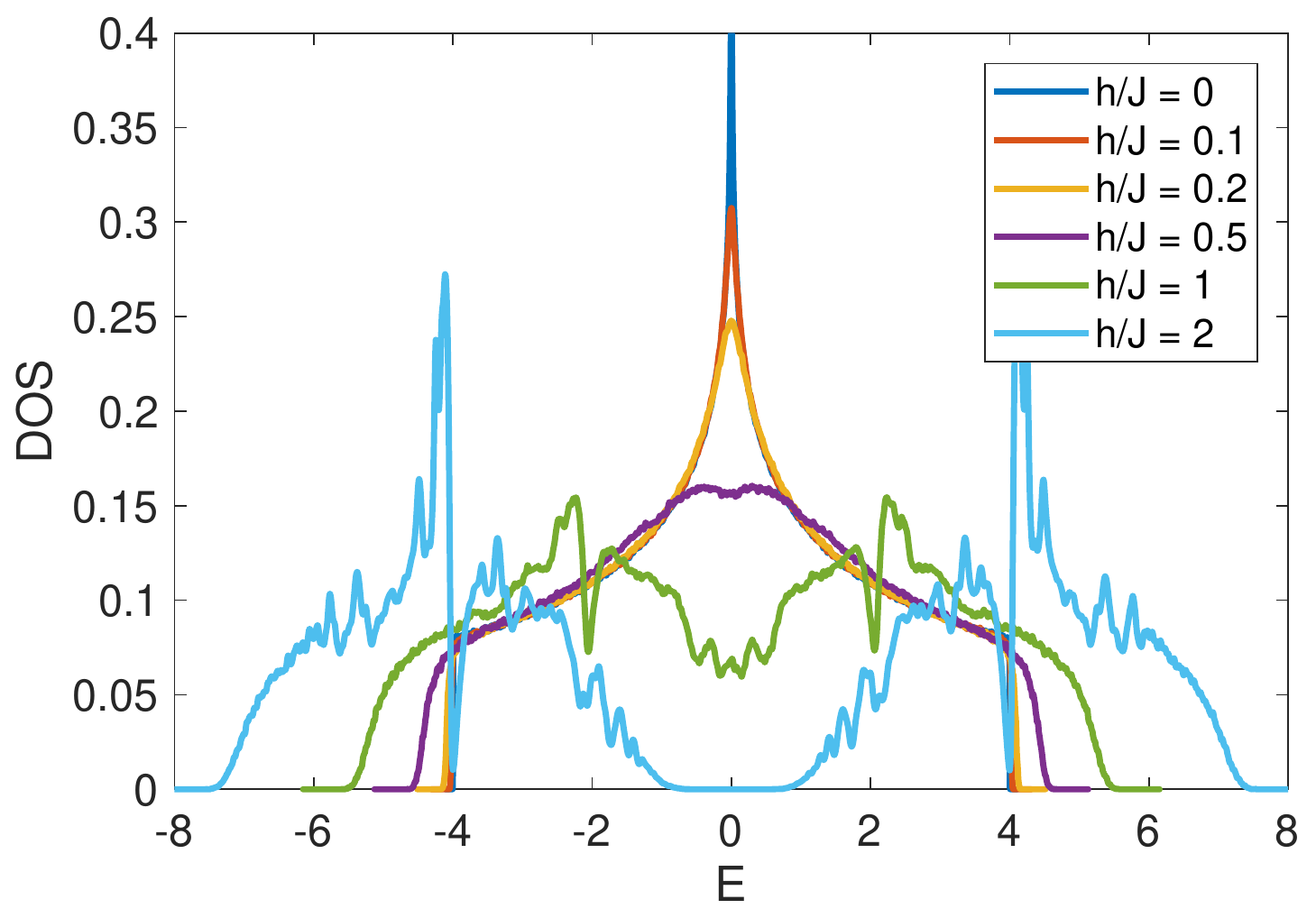}
	\subfigimg[width=.44\textwidth]{\hspace*{0pt} \textbf{(b)}}{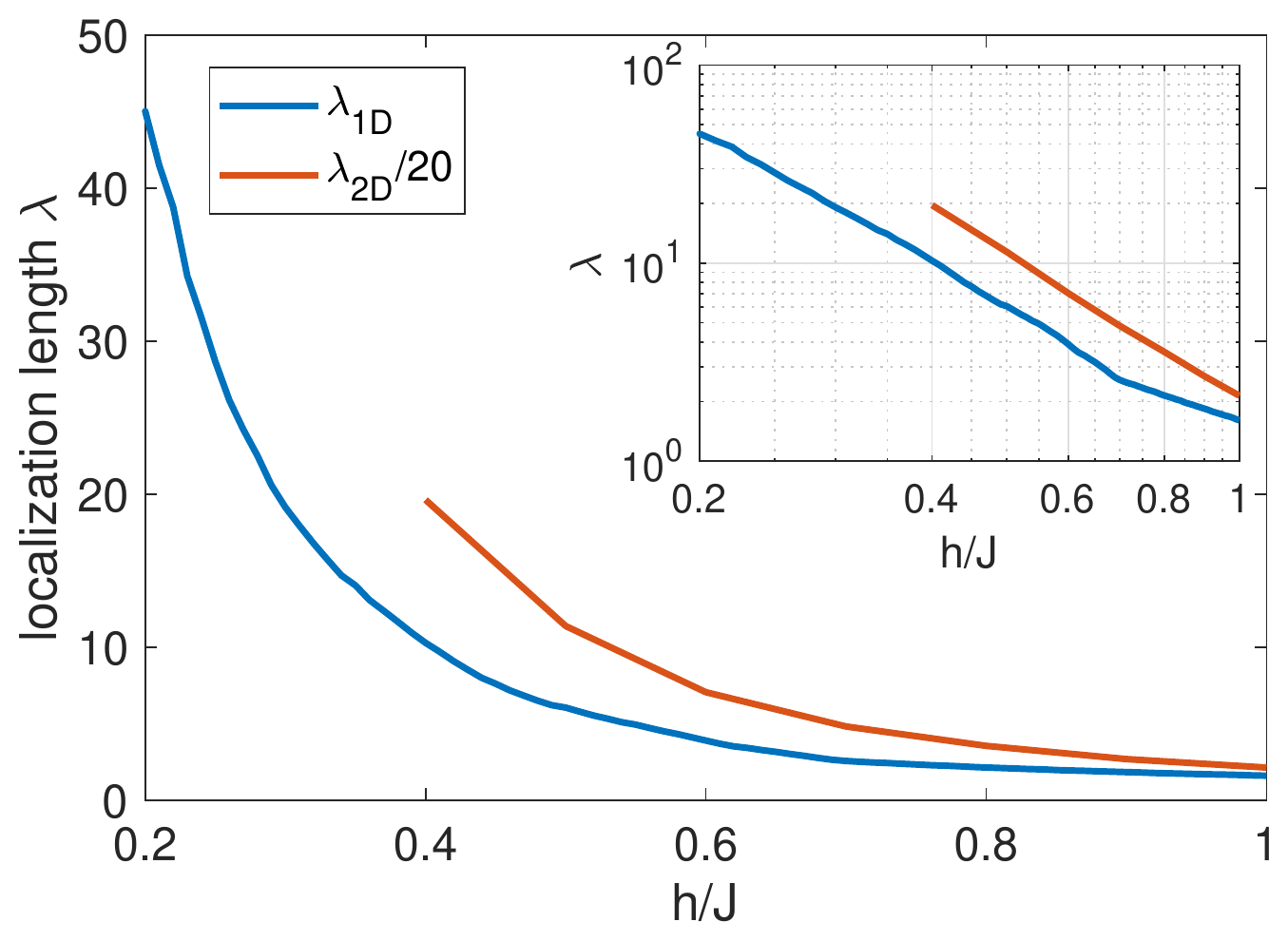}
	\caption{(a) Single-particle density of states for a 2D square lattice for different $h/J$. The DOS is computed using the kernel polynomial method, see Appendix~\ref{ap: KPM}. (b) Localization length in 1D and 2D. In 1D, localization length is computed using the spectral formula~\eqref{eq: spectral locLength} as explained in the main text. In 2D, the localization length is scaled by a factor of $20$ and is computed by the transfer matrix method on a strip of width $100$ sites and length 250,000 sites.}\label{fig: 2D DOS}
\end{figure}

For all diagnostics we consider a cut through the system e.g.~perpendicular to the domain wall. As in 1D, we find that the density imbalance saturates at a non-zero value at long times, see Fig.~\ref{fig: 2D results}(a). However, in contrast to 1D, the localization length is larger in 2D (for the same $h/J$) leading to smaller long-time values for $\Delta\rho$ and larger values for $N_\text{half}$. Furthermore, for the values of $h/J$ shown in Fig.~\ref{fig: 2D results} which are much larger than those presented for the 1D case, the amplitude of the fluctuations is much smaller. In other words, the extra dimension produces a damping effect on these fluctuations because of the much smoother single-particle DOS, even for $h/J>1$, see Fig.~\ref{fig: 2D DOS}(a). We also find that for the checkerboard initial state, the density imbalance persists more than for the charge density wave, as shown in the inset. This can be understood by considering the action of the Hamiltonian on this initial state. In contrast to the CDW, in the checkerboard case, delocalization of fermions appears at higher order in $J/h$, thus constraining the fermion dynamics in the checkerboard initial state compared to a charge density wave. Comparison of the corresponding 1D and 2D results shows that the remaining imbalance is generally much smaller in 2D than in 1D, which is due to the fact that localization lengths are much larger in 2D, see Fig.~\ref{fig: 2D DOS}(b).

Starting from the domain wall initial states, we can again see a linear initial spreading which is halted due to the effective disorder, see Fig.~\ref{fig: 2D results}(b-c). In this case we do not find long-lived oscillations for $h/J > 1$. Our results clearly show that the localization length is much larger in 2D compared to 1D, which can be seen from the dynamics of domain wall spreading (on a larger lengthscale) for the same disorder strength as in 1D. We can use again  $N_\text{half}$ to quantify the dynamics of domain wall spreading. This observable approaches a finite value in the localized case, see Fig.~\ref{fig: 2D results}(d).

If we compare the DOS for 2D with that of 1D we notice some important similarities and differences. First, we see a gap opening as in 1D for large values of $h$. Whereas in the 1D case this gap appears at $h=J$, the spectrum is still gapless for $h$ order of $J$ in 2D. There is also an increase in the bandwidth, both of the total DOS and of the individual sub-bands that develop in the large $h$ limit, owing to the extra dimension. More importantly we find that sub-bands remain much smoother than in 1D for a much wider range of effective disorder strength.

In Fig.~\ref{fig: 2D DOS} we show the dependence of the maximum localization length on $h/J$ in 1D and 2D. The localization length is the characteristic length scale of the exponential tails of the single-particle wavefunctions defined via $e^{-j/\lambda}$. In the 1D case the results are obtained using the spectral formula~\cite{Kramer1993,Thouless1972}
\begin{equation}\label{eq: spectral locLength}
\frac{1}{\lambda} = \min_{E} \int^\infty_{-\infty} g(x) \ln|E-x| \;\text{d}x,
\end{equation}
where $g(x)$ is the DOS calculated via the kernel polynomial method (Appendix~\ref{ap: KPM}). In the 2D case we used the transfer matrix method~\cite{Kramer1993}, see Appendix~\ref{ap: transfer matrix}. The 2D results are rescaled by a factor of 20 which demonstrates an order of magnitude difference in localization lengths in 1D and 2D. However, the localization length as a function of disorder strength shows similar power-law in 1D as in 2D.

\begin{figure}[t]
	\centering
	\subfigimg[width=.15\textwidth]{\hspace*{-10pt} \textbf{(a)}}{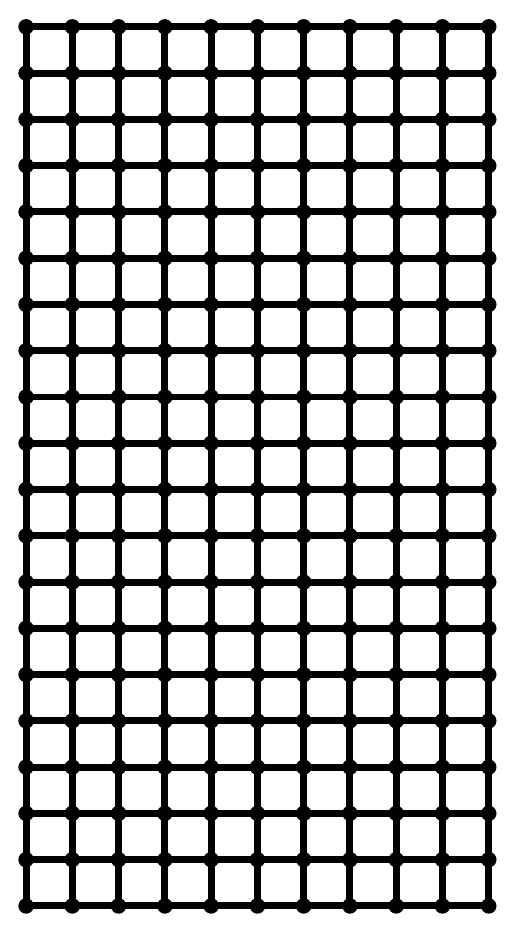}
	\;\;\subfigimg[width=.15\textwidth]{\hspace*{-10pt} \textbf{(b)}}{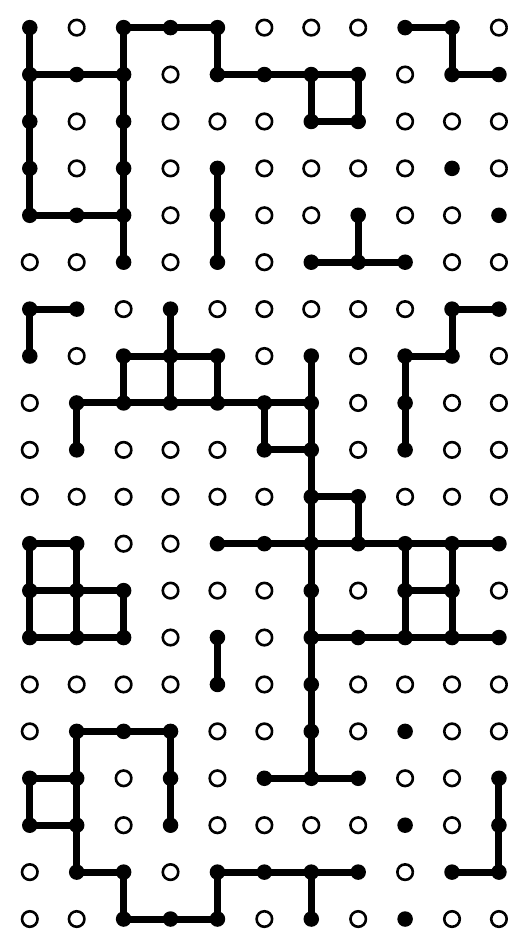}
	\;\;\subfigimg[width=.15\textwidth]{\hspace*{-10pt} \textbf{(c)}}{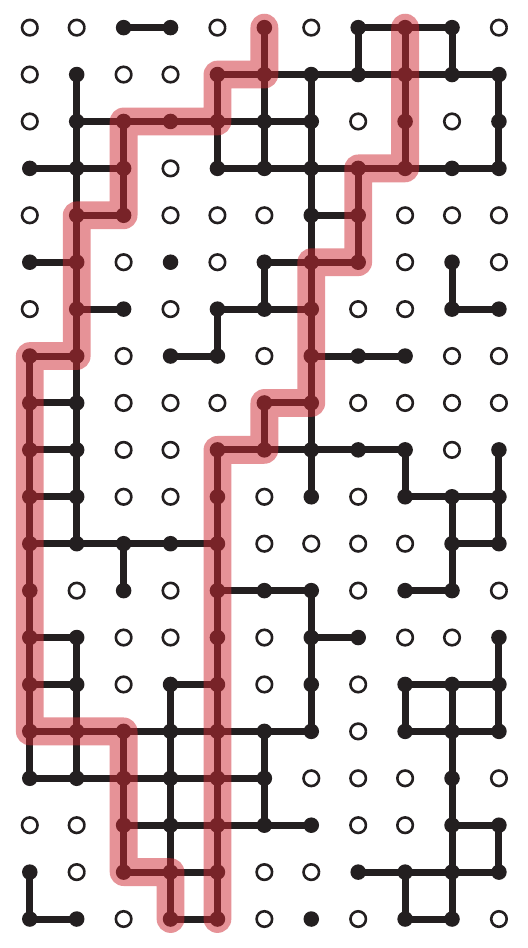}
	\caption{Schematic picture of the site percolation problem. (a) A fully connected lattice on which fermions can hop in the limit $h/J=0$. (b-c) When $h/J\gg 1$ the fermions are constrained to hopping between sites with the same effective potential (filled sites and bonds) and the other sites become inaccessible (open circles) leading to a quantum site percolation problem. (b) connected sites for the bias $p=0.5$ showing that the absence of paths (in the thermodynamic limit) which connect opposite sides of the lattice. (c) the same for the bias $p=0.7$ which is above the percolation threshold $p_c\approx 0.5927$, showing in red connected paths across the system.}
	\label{fig: percolation}
\end{figure}

\begin{figure}[th!]
\centering
\subfigimg[width=.45\textwidth]{\hspace*{0pt} \textbf{(a)}}{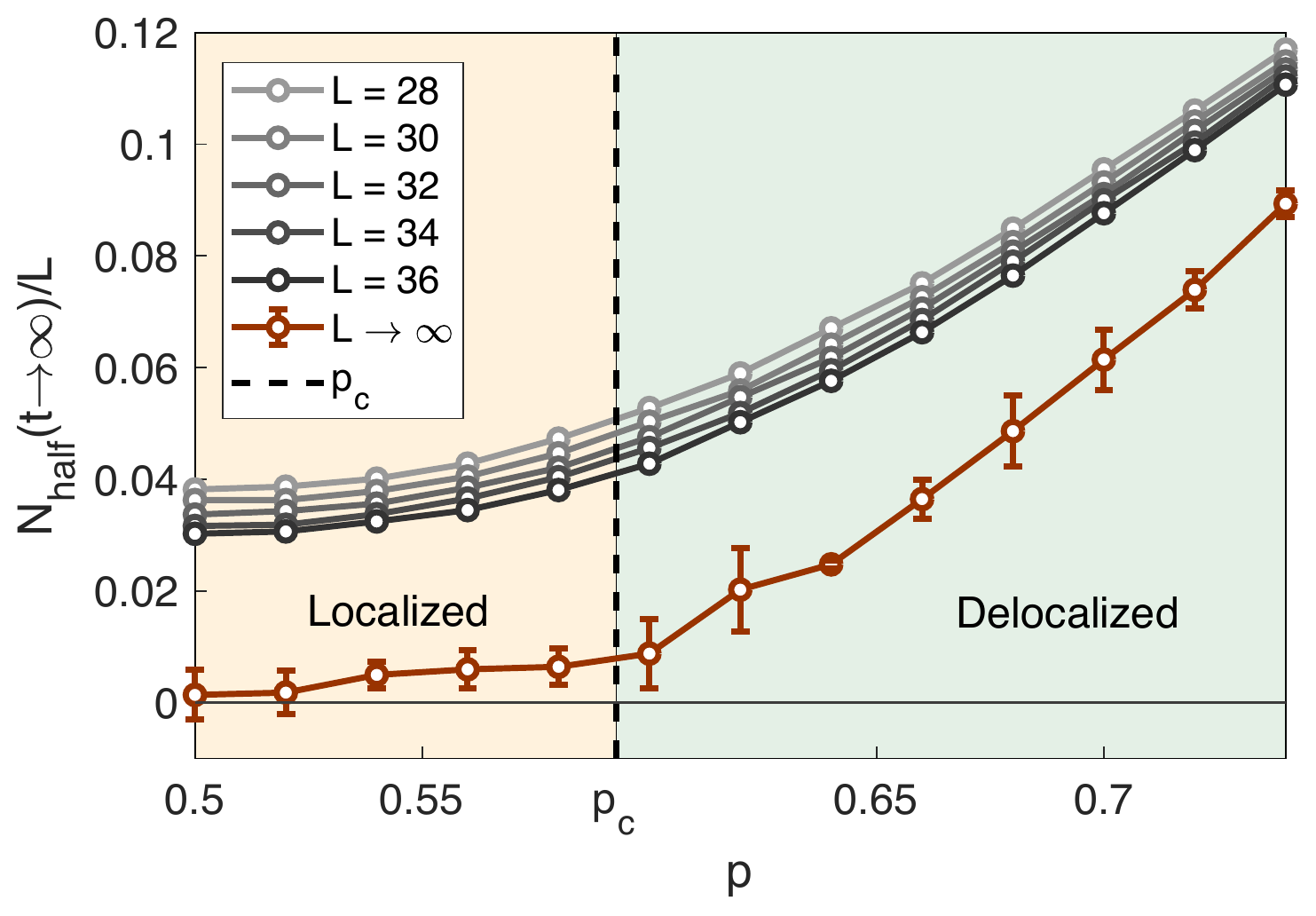}
\subfigimg[width=.45\textwidth]{\hspace*{0pt} \textbf{(b)}}{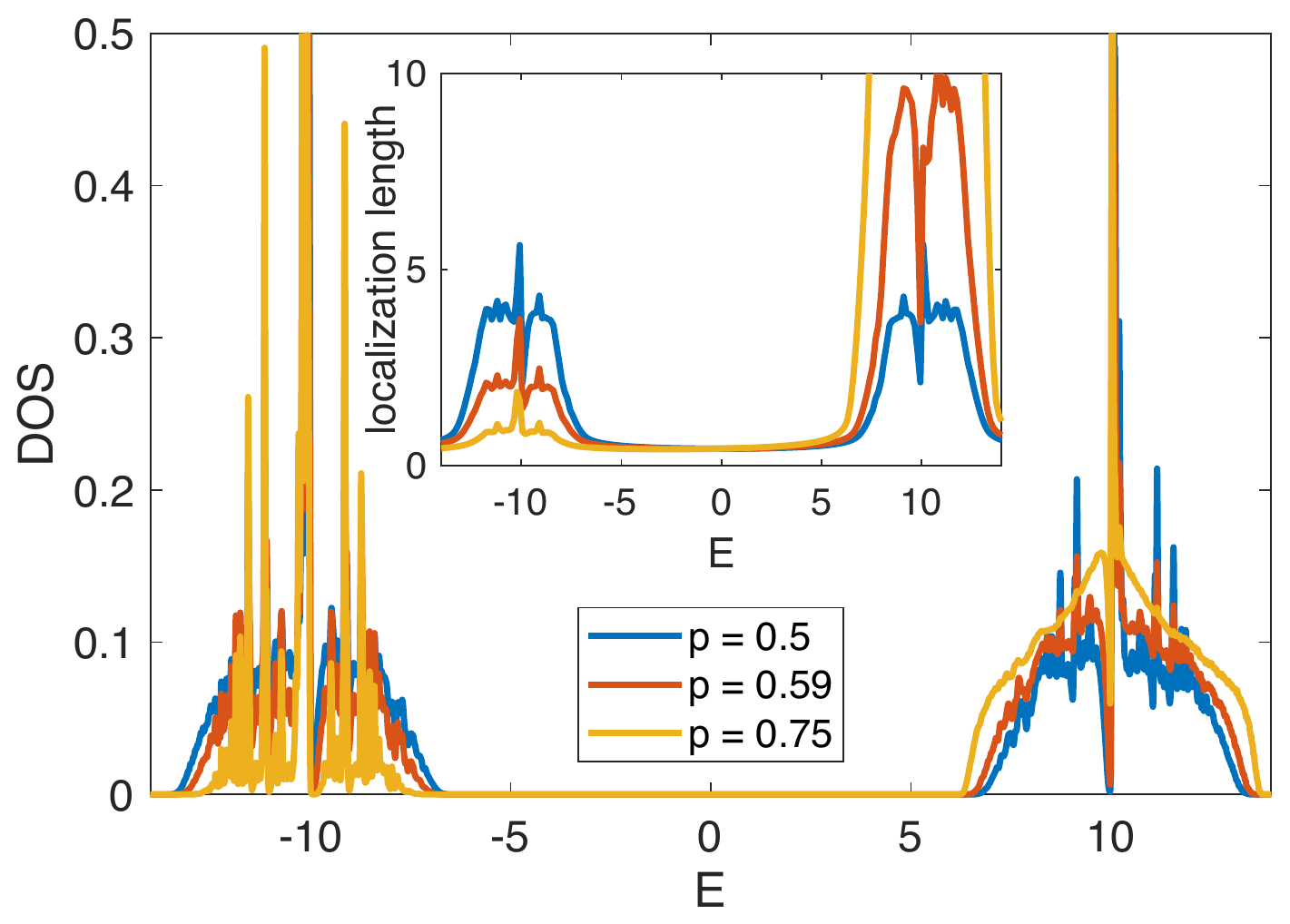}
\caption{(a) Total number of particles making it across a domain wall $N_\text{half}$ for various values of the charge distribution bias, $p$. The figure shows the rescaled value $N_\text{half}/L$, and the extrapolated limit $L\rightarrow\infty$, where $L$ is the linear dimension of the system. Error bars correspond to 2 standard deviations in the linear fit from finite size scaling. Note that $p_c\approx0.5927$ corresponds to the classical site percolation threshold for a square lattice. (b) Density of states for different values of the bias $p$ and $h/J=20$. Inset shows the corresponding energy-resolved localization length. The results shown in (a) were obtained using the determinant method (Appendix~\ref{ap: determinant}). We used the KPM (Appendix~\ref{ap: KPM}) for the DOS in (b) and the transfer matrix approach for the localization length (Appendix~\ref{ap: transfer matrix}) for a system of $150\times250000$ sites.}\label{fig: Nhalf bias}
\end{figure}

\subsection{Delocalization in 2D and quantum percolation}\label{sec: delocalization}

It is known that all single-particle states in 1D and 2D are localized in the presence of disorder. However, it is possible to have delocalized states and even a mobility edge separating localized and delocalized states in case of correlated disorder. The famous example in 1D is the Aubry-Andre model which has a periodic potential incommensurate with the lattice. Another example is when time-reversal symmetry is broken, for example by a magnetic field.

Due to the binary nature of the disorder potential in our model, we can get delocalized states in 2D without the need of correlated disorder. This can be achieved by biasing the distribution of charges $q$ with a probability $p$, such that $q = \pm 1$ with probability $p$ and $1-p$, respectively. Alternatively, one could impose a stricter global constraint $N^{-1}\sum_i (\hat{q}_i + 1)|\Psi\ra = p |\Psi\ra$. In the large $h$ limit of strong disorder, we arrive at a quantum site percolation problem~\cite{Alvermann2005}. The lattice then is decomposed into parts with sites sitting at the top or bottom of the binary potential, and to the lowest order only hopping between sites with the same sign of potential is allowed. This can be appreciated by considering Eq.~\eqref{eq: strong h Hamiltonian} on a 2D lattice. In $\hat{H}_h$, there will only be hopping terms between neighbouring sites with the same values of $q$, see Fig.~\ref{fig: percolation}. Since the threshold in the classical site percolation problem is $p_c \approx 0.5927$, this is consistent with having localized wavefunctions for all $h$ for our typical setting of $p = 1/2$. However, if we set $p > p_c$, or $1-p > p_c$, then we should expect percolation in our model for large $h$. Note that if we get percolation, say at the top of the potential, then we necessarily localize the modes at the bottom of the potential more, and vice versa.

In order to understand the effect of percolation, we study the time evolution from a domain wall initial state with changing system size and bias $p$, as shown in Fig.~\ref{fig: Nhalf bias}(a). We plot $N_\text{half}(t\rightarrow\infty)/L$, where $L$ is the linear dimension of a square lattice with $N = L\times L$ sites. If particles are localized then we would expect $N_\text{half}(t\rightarrow\infty)/L$ to tend to zero as $L$ is increased -- i.e., $N_\text{half}$ is finite and independent of $L$. Whereas, if the particles are delocalized, we should find that a finite proportion of the particles makes it across the domain wall, $N_\text{half}(t\rightarrow\infty)/L \rightarrow \text{constant}$. In Fig.~\ref{fig: Nhalf bias}(a) we show the extrapolation from finite size scaling, which agrees with this expected behaviour in the two limits. Furthermore, the change in behaviour is observed to happen around the critical percolation threshold $p = p_c$.

We can also understand this percolation behaviour by looking at the DOS and the energy-resolved localization length as a function of $p$, as shown in Fig.~\ref{fig: Nhalf bias}(b). As $p$ is increased past the critical point we see a clear asymmetry with respect to energy in these results. In the DOS, one of the sub-bands becomes similar to that of the large $h$ limit of $p=1/2$ problem. The other sub-band becomes much smoother and similar to the DOS for a clean 2D system. Furthermore, we respectively see a decrease and increase in the localization length in the these two sub-bands, see inset, which is consistent with percolation of the fermions at a positive potential and the localization of those at a negative potential.

Because we are studying a quantum model, it is not clear that there should be a direct correspondence with site percolation. In particular, given a path through the system, as in Fig.~\ref{fig: percolation}, we would generally expect quantum fluctuations to lead to backscattering, which may hinder conductance. References~\cite{Schubert2008,Schubert2009} showed, using large scale numerics with the kernel polynomial methods, that the quantum percolation threshold $p_c^q < 1$, and we necessarily have $p_c\leq p_c^q$. Furthermore, studies on the Bethe lattice seem to show that the quantum site percolation threshold agrees with the classical threshold~\cite{Alvermann2005}, and our results are also consistent with the classical site percolation on a square lattice. However, this point is far from having been settled and   Ref.~\cite{Schubert2009} observes that $p_c < p_c^q$ by doing a careful analysis using local density of states for much larger systems. In particular they find that $p_c^q > 0.65$, which we do not see in our results. However, because of the modest system sizes used in our calculations, our results may still show finite-size effects. 

\subsection{Localization in 3D}\label{sec: delocalization 3D}

For a 3D cubic lattice it is generally believed that $p_c < p_c^q < 1$, though still $p_c^q < 0.5$~\cite{Schubert2009}. Therefore a 3D case offers an interesting setting for studying localization. In 3D, there is a critical disorder strength needed to achieve localization. However, in the large $h/J$ limit we would expect delocalized states for all values of the bias probability $p$. One can expect delocalized states for both low and high $h/J$ but localized states for intermediate values. 

\section{Localization in presence of interactions}\label{sec: interactions}

Our model can be modified in a variety of ways to include interactions. Here we consider a subset of such extensions, focussing on the 1D case because of numerical limitations. Terms which can be added to the Hamiltonian~\eqref{eq: H general} fall into two classes depending on whether they give dynamics to conserved charges. Note that these terms in general lead to interactions in contrast to the Hamiltonian~\eqref{eq: H qs}. In the presence of interactions, it is not possible to use determinant methods, and instead we have to resort to exact diagonalization, and on Krylov subspace methods to calculate the time evolution, see Appendix~\ref{ap: Krylov}. 

\subsection{Conserved charges and many-body localization}\label{sec: conserved}

Those terms that commute with the charges include fermion density-density interactions and  longitudinal field terms:
\begin{equation}\label{eq: additional commuting}
\Delta \sum_{\la j k \ra } \hat{n}_{j} \hat{n}_k, \quad \text{and} \quad B_x \sum_j \hat{\sigma}^x_{jk},
\end{equation}
where $\hat{n}_j = \hat{f}^\dagger_j \hat{f}_j$.

Commutation relations between these terms and the charges are most easily seen when they are expressed in terms of the original degrees of freedom, $\hat{q}_j = \hat{\sigma}^x_{j-1,j} \hat{\sigma}^x_{j,j+1} (-1)^{\hat{n}_{j}}$. The charges clearly do not commute with any local terms containing $\hat{\sigma}^z_j, \hat{\sigma}^y_j$, but do commute with a longitudinal magnetic field. Electron density-density interactions retain their form under the transformation from $f$ to $c$ fermions, and the density operator also commutes with the fermion parity operator appearing in the expression for conserved charges.

First, we consider adding nearest-neighbour density-density interactions between fermions. Up to constant terms the model can be written as
\begin{equation}\label{eq: H MBL}
\begin{aligned}
\hat{H} = -J \sum_{\la j k \ra} \hat{\sigma}^z_{jk} \hat{f}^\dag_j \hat{f}_k &- h \sum_j \hat{\sigma}^x_{j-1,j} \hat{\sigma}^x_{j,j+1}\\ &+ \Delta \sum_j (2\hat{n}_j -1) (2\hat{n}_{j+1} - 1).
\end{aligned}
\end{equation}
Under transformation to $c$-fermions and charges, the first two terms transform as before and the fermion interactions remain invariant since $\hat{f}^\dag_j \hat{f}_j \equiv \hat{c}^\dag_j \hat{c}_j$. We can then use a Jordan-Wigner transformation, $\hat{S}^+_j = \hat{c}^\dag_j (-1)^{\sum_{l<j} \hat{n}_l}$, and $\hat{S}^z_j = \hat{n}_j = \frac{1}{2}$ to cast the Hamiltonian in the following form
\begin{equation}
\begin{aligned}
\hat{H}_{XXZ} = - J \sum_{j} (&\hat{S}^+_j \hat{S}^-_{j+1} + \hat{S}^-_j \hat{S}^+_{j+1} ) \\ &+ 4\Delta \sum_j \hat{S}^z_j \hat{S}^z_{j+1} +2h \sum_j \hat{q}_j \hat{S}^z_j,
\end{aligned}
\end{equation}
which is an XXZ Hamiltonian describing a spin chain with a binary potential set by $q_j = \pm 1$. This XXZ Hamiltonian with quenched disorder serves as one of the paradigmatic models of many-body localization. Although in the context of MBL this model is usually studied with uniformly sampled disorder~\cite{Znidaric2008,Bardason2012}, is has also been studied in the case of binary disorder~\cite{Andraschko2014,Tang2015,Enss2016}. Using this mapping, and looking at dynamical correlators we observe the behaviour usually found in MBL phases. However, note that our starting point is a disorder-free Hamiltonian~\eqref{eq: H MBL}.

\begin{figure}[b]
	\centering
	\!\!\!\!\subfigimg[width=.46\textwidth]{\hspace*{0pt} \textbf{(a)}}{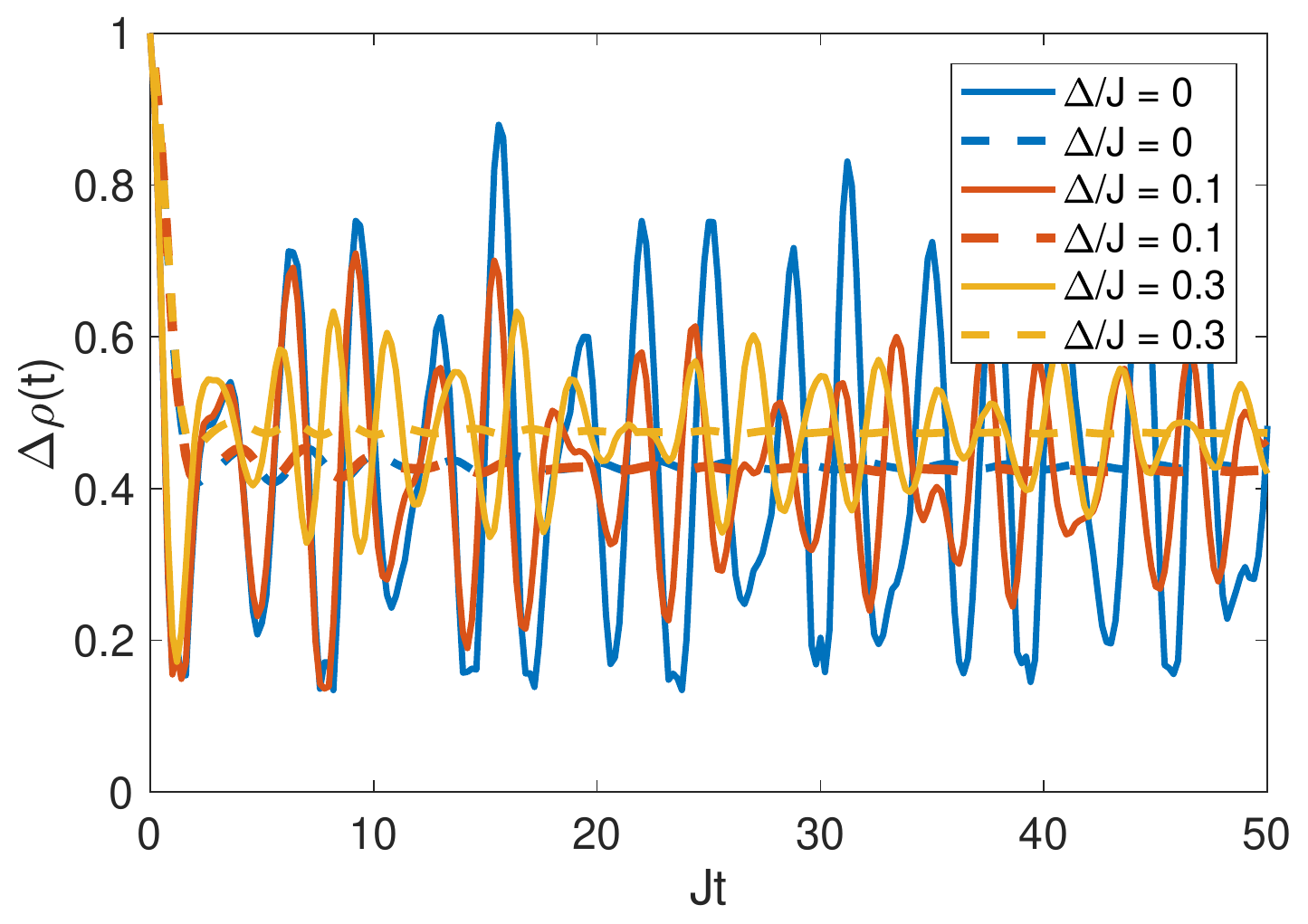}
	\subfigimg[width=.45\textwidth]{\hspace*{0pt} \textbf{(b)}}{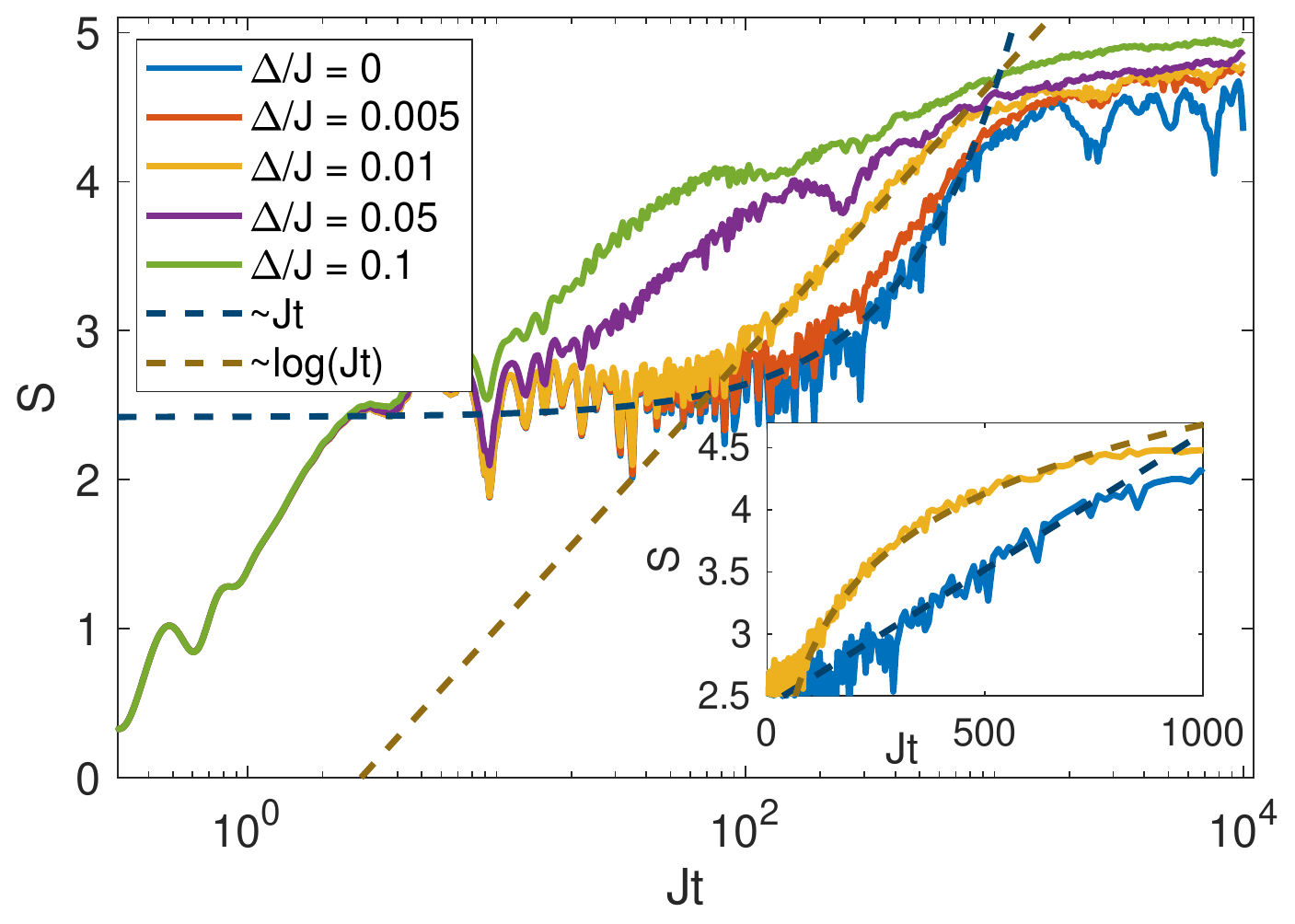}
	\caption{Quantum quench from an initial charge density wave state, $h/J=20$. (a) Density imbalance $\Delta\rho(t)\propto \sum_j |\la 0 | \hat{n}_j(t) - \hat{n}_{j+1}(t) |0\ra|$ and the time-averaged value of $\frac{1}{t}\int^t_0\text{d}\tau\,\Delta\rho(\tau)$ (dashed lines) after the same quench.
		(b) Von Neumann entanglement entropy computed using ED for $N=12$ sites (thin, light) for various values of $\Delta$ shown on a semi-log scale. The spatial bipartition is taken along the central bond. (inset) The same data on a linear scale for $\Delta/J = 0, 0.01$. Dashed lines show fitted linear and logarithmic curves.}
	\label{fig: MBL}
\end{figure}

Let us consider the charge density wave initial state as explained in the non-interacting case, see Fig.~\ref{fig: MBL}(a). We find that the density imbalance $\Delta\rho$ saturates at a non-zero value (as in the non-interacting case) indicating the persistent memory due to localization. For small interactions, the asymptotic value is close to the one for the non-interacting case for $\Delta = 0.1J$, but as the interaction strength is increased it also acts to stabilise the charge density wave which leads to an increase of the asymptotic value, as can already be seen for $\Delta = 0.3J$. We also observe that the interactions have the effect of damping the fluctuations around this asymptotic value, which is evident even on the short time scales shown in Fig.~\ref{fig: MBL}(a). 

Next, we can consider the von Neumann entanglement entropy, shown in Fig.~\ref{fig: MBL}(b). Here we see a qualitative change (compared to the non-interacting case) in the entanglement growth following the initial area-law plateau. While in the non-interacting case we get linear growth followed by saturation, in presence of density-density interactions, we observe a slower logarithmic growth, as shown by the dashed lines in Fig.~\ref{fig: MBL}(b), and in the inset. This logarithmic behaviour, which sets in at times $\sim\Delta/J$, is consistent with the phenomenology of MBL.

Let us now consider another term which can be added to the Hamiltonian, and which commutes with the charges, namely the longitudinal field. The latter no longer commutes with plaquette operators and thus the duality mapping~\eqref{eq: duality} is no longer useful. However, we still have the conserved quantities $\hat{q}_j = \hat{\sigma}^x_{j-1,j} \hat{\sigma}^x_{j,j+1} (-1)^{\hat{n}_j}$. In the original spin picture, the effect of this term is to confine spin excitations~\cite{Kormos2016}. 

\begin{figure}[t]
	\centering
	\subfigimg[width=.45\textwidth]{\hspace*{0pt} \textbf{(a)}}{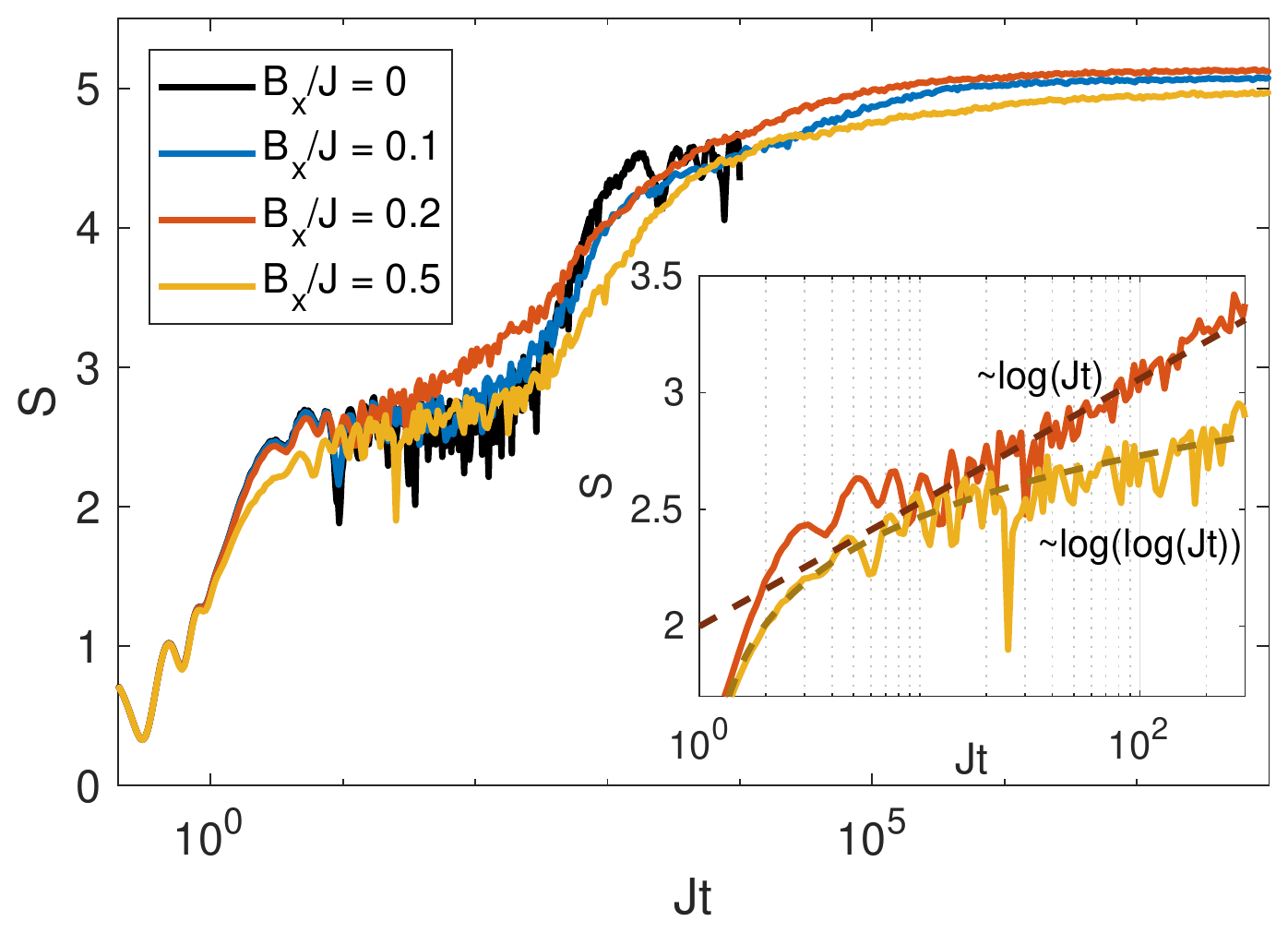}
	\subfigimg[width=.45\textwidth]{\hspace*{0pt} \textbf{(b)}}{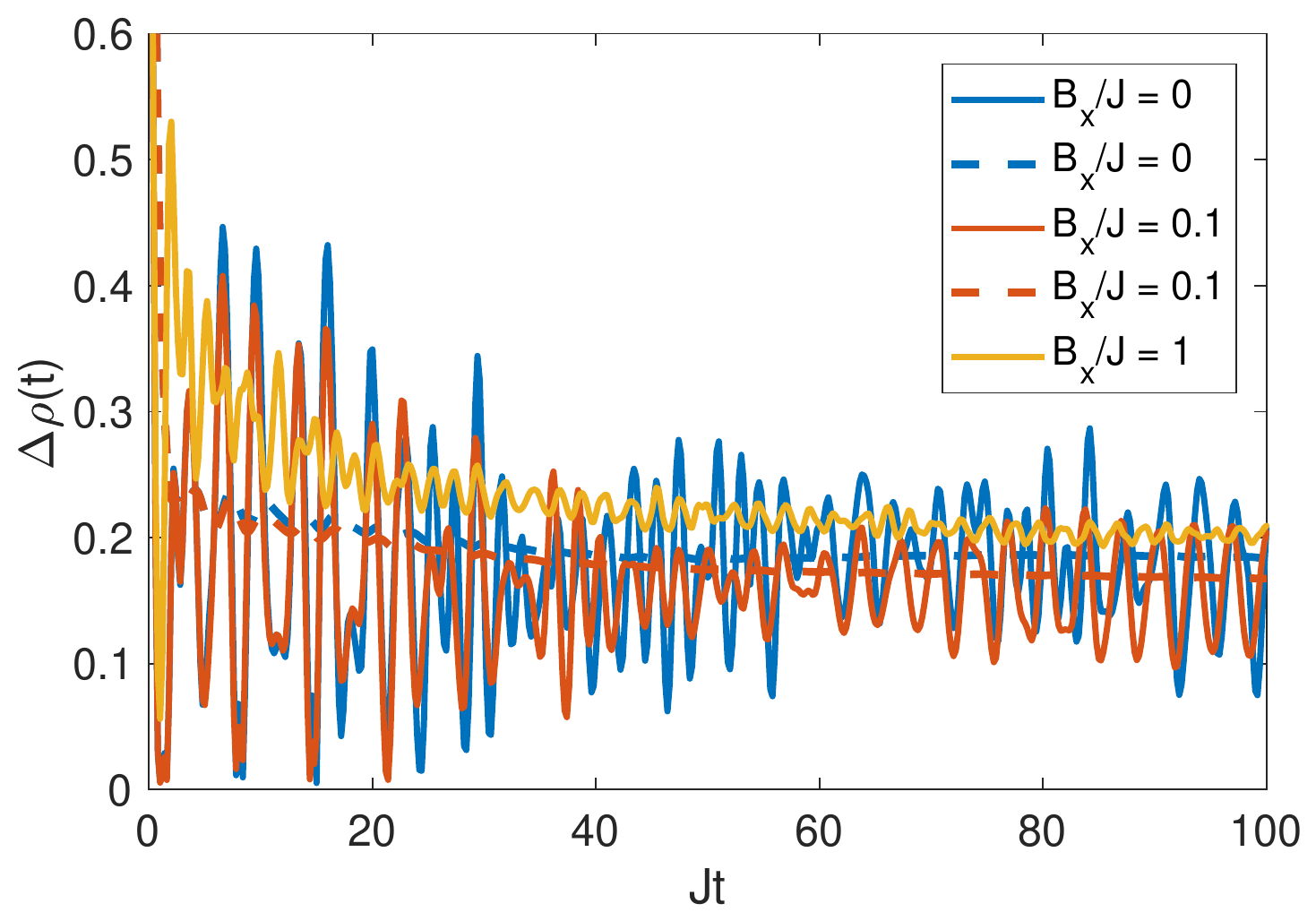}
	\caption{(a) Entanglement entropy after a quantum quench for various values of the longitudinal field $B_x$ shown on a semi-log scale. A window of the same data for $B_x/J = 0.2,0.5$ is given in the inset. Dashed lines correspond to $\ln(t)$ and $\ln(\ln(t))$ behaviour. (b) Density imbalance after a quench from a charge density wave. Results obtained using ED.}
	\label{fig: longitudinal}
\end{figure}

In Fig.~\ref{fig: longitudinal}(a) we present the results for entanglement entropy with the Hamiltonian having a longitudinal field term whose strength is controlled by $B_x$. The results show a rich behaviour. In particular, one can notice two new qualitative features. For small $B_x/J\sim0.2$, we observe logarithmic entanglement growth at a time scale set by $~B_x/J$ similar to the MBL behaviour observed above. However, for larger $B_x/J\sim0.5$ we find a slower growth, which can be fit with $\ln(\ln(t))$ as shown in the inset. Furthermore, for small $h$, the interactions generate extra entanglement compared to the non-interacting results, whereas for larger $B_x$ the entanglement is reduced. We also see this behaviour in the density imbalance starting from a charge density wave, shown in Fig.~\ref{fig: longitudinal}(b). We again see that interactions have a damping effect on the fluctuations and that a strong enough field stabilises the charge density wave.

\begin{figure*}[!tb]
	\centering
	\subfigimg[width=.33\textwidth,valign=b]{\hspace*{0pt} \textbf{(a)}}{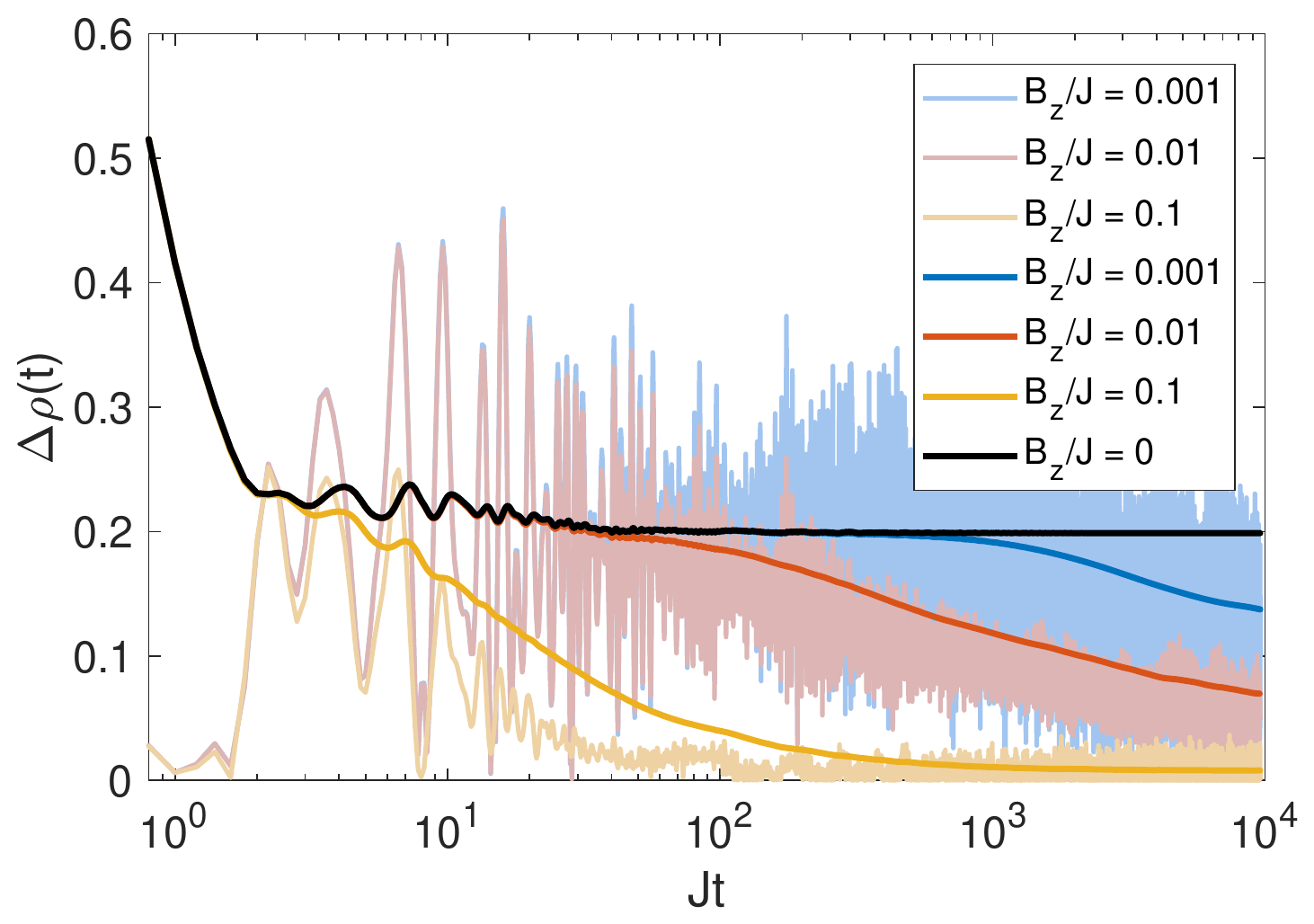}
	\subfigimg[width=.33\textwidth,valign=b]{\hspace*{0pt} \textbf{(b)}}{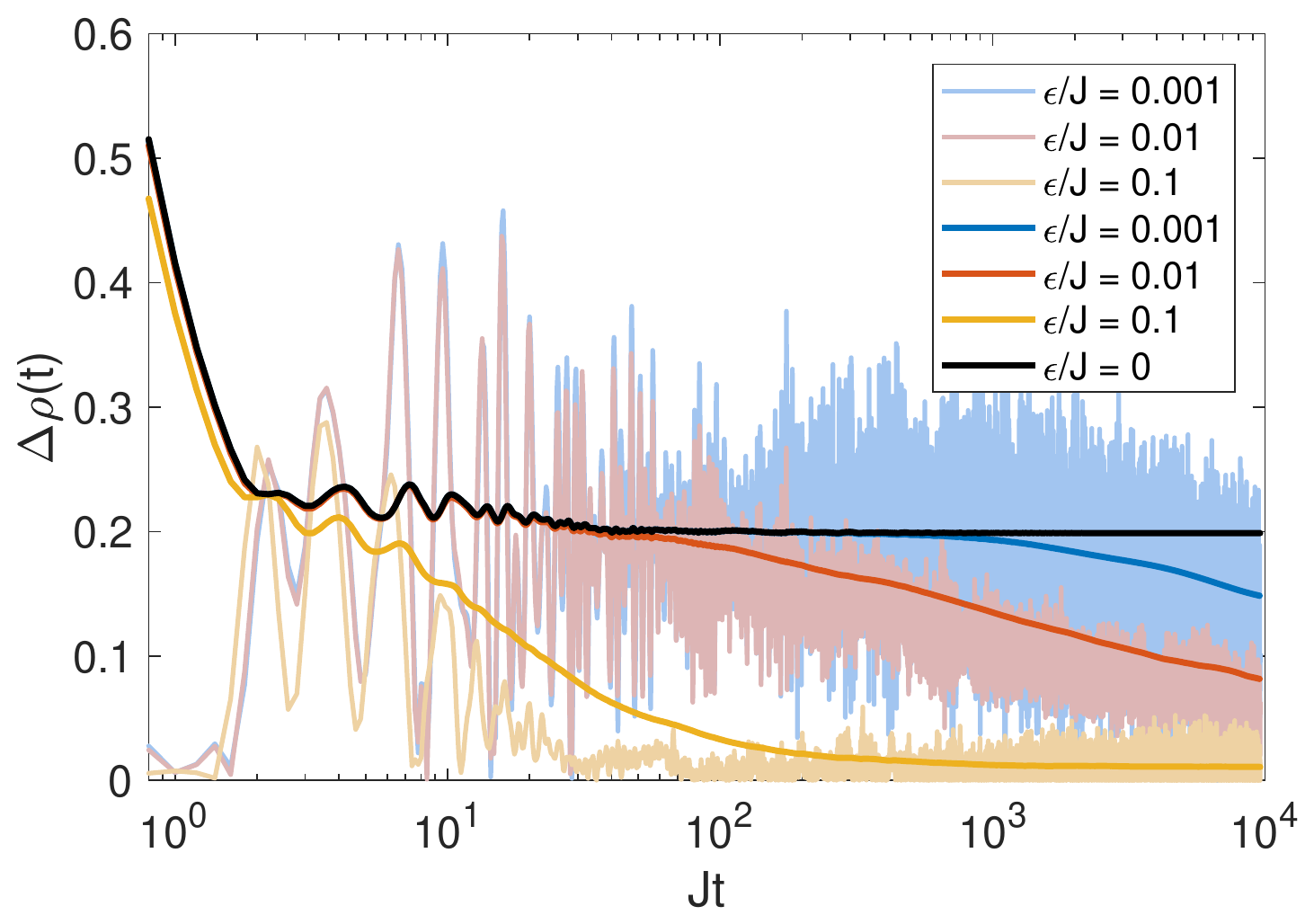}
	\subfigimg[width=.33\textwidth,valign=b]{\hspace*{0pt} \textbf{(c)}}{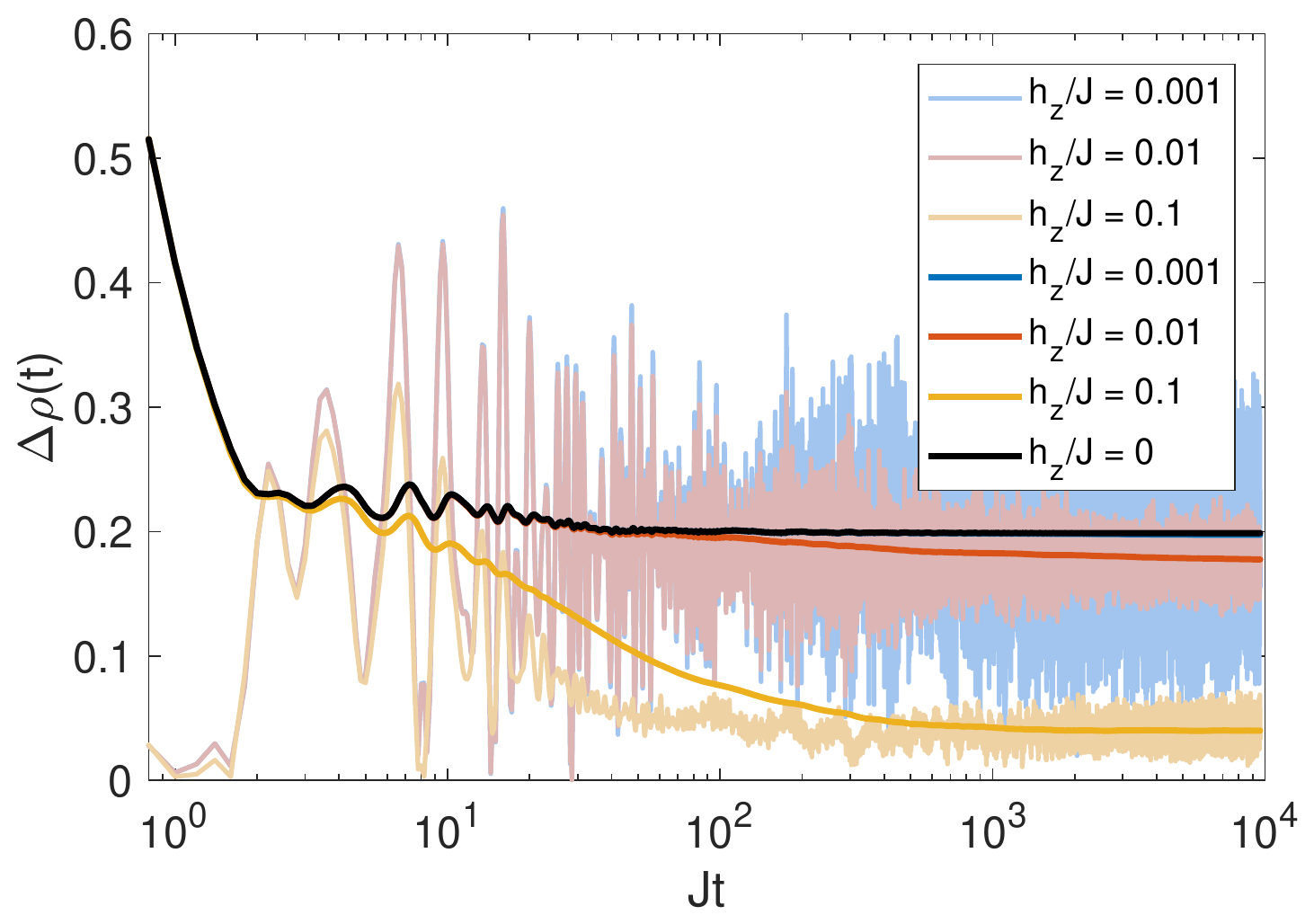}
	\caption{Density imbalance after a quantum quench from a CDW with dynamical charges. We study a system with is $N=10$ sites with periodic boundary conditions and $h=J$. The light curves are the imbalance $\Delta\rho$ and the dark thick lines are the time averaged value $\frac{1}{t}\int^t_0 \text{d}\tau \Delta\rho(\tau)$. (a) With an additional transverse field $B_z\sum_i\hat{\sigma}^z_{i,i+1}$. (b) Additional fermion hopping $\epsilon\sum_{\la i j\ra} \hat{f}^\dagger_i\hat{f}_j$. (c) Added transverse Ising coupling $h_z \sum_{i} \hat{\sigma}^z_{j-1,j} \hat{\sigma}^z_{j,j+1}$. Dynamics is computed using Krylov subspace methods (Appendix~\ref{ap: Krylov}).}\label{fig: Int Breaking}
\end{figure*}

\begin{figure}[!tb]
	\centering
	\includegraphics[width=.45\textwidth]{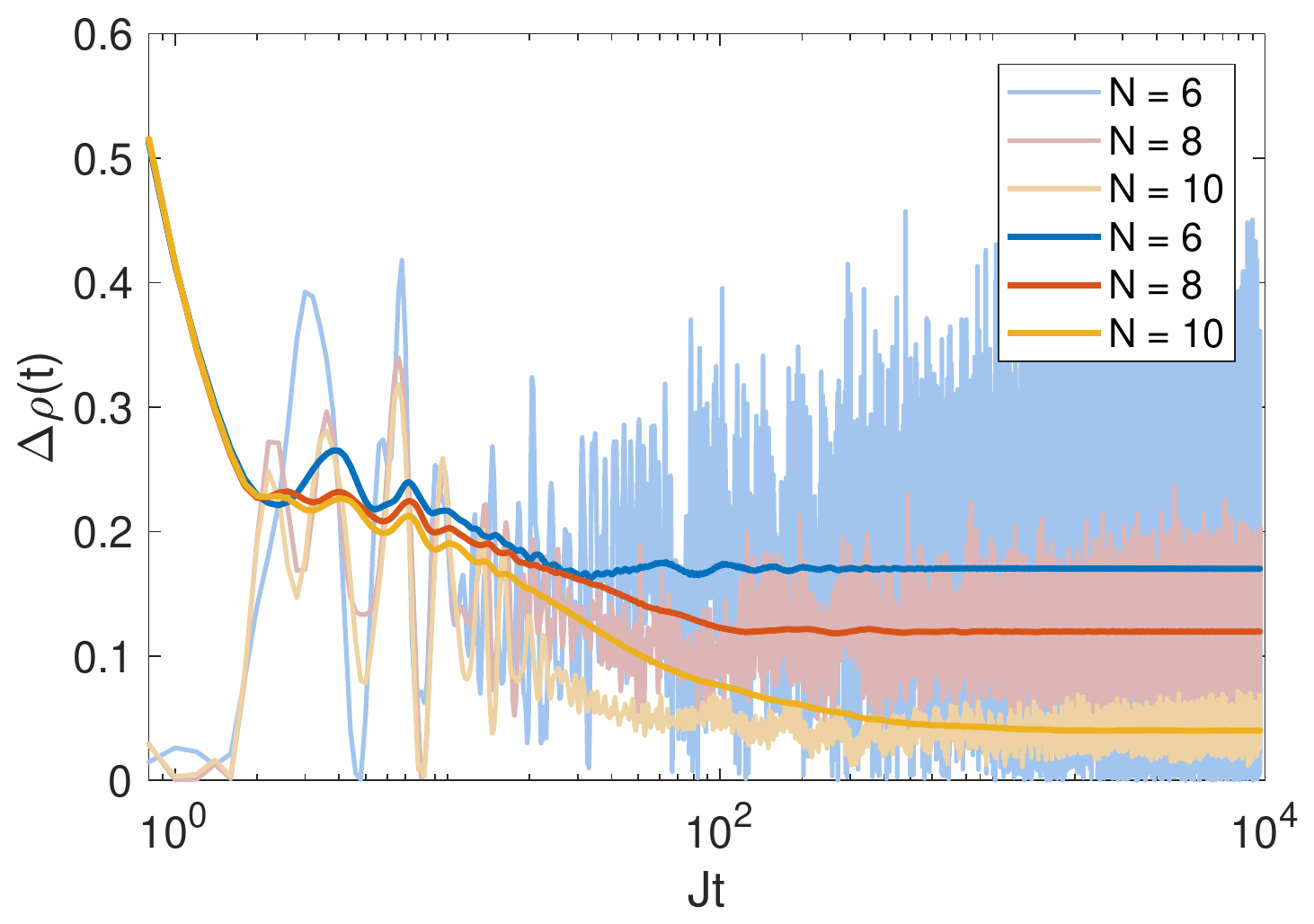}
	\caption{Finite size scaling of the asymptotic density imbalance in presence of additional Ising coupling $h_z \sum_{i} \hat{\sigma}^z_{j-1,j} \hat{\sigma}^z_{j,j+1}$. We use system sizes $N=6,8,10$ with $h_z = 0.1J$ and $h=J$. Light curves show the density imbalance $\Delta\rho$ and the dark and thick curves show the time averaged value of $\frac{1}{t}\int^t_0 \text{d}\tau \Delta\rho(\tau)$. See Appendix~\ref{ap: Krylov} for details of the Krylov subspace numerical method used.}\label{fig: Int Breaking Scaling}
\end{figure}

\subsection{Dynamical charges and quasi-MBL}\label{sec: quasi-MBL}

In the second category of terms, i.e. those terms that do not commute with charges and generate their dynamics, we consider three types of terms:
\begin{equation}
h_z \sum_j \hat{\sigma}^z_{j-1,j} \hat{\sigma}^z_{j,j+1}, \quad B_z \sum_j \hat{\sigma}^z_{j,j+1}, \quad \epsilon \sum_{\la i j \ra} \hat{f}^\dag_i \hat{f}_j.
\end{equation}
The localization behaviour discussed in above relies on the presence of static charges $\hat{q}_j$, which act as an effective disorder. It is therefore a natural question to ask what happens when we give dynamics to the charges.

Fig.~\ref{fig: Int Breaking} shows the effect of these additional interactions on the density imbalance after a quench from the charge density wave initial state. Figs.~\ref{fig: Int Breaking}(a-b) clearly show that the introduction of $B_z$ and $\epsilon$ leads to the decay of this imbalance, and ultimately to the disappearance of the charge density wave. One can also see that the time scale at which the results significantly deviate from the $B_z=\epsilon=0$ case is determined by $B_z^{-1}$ and $\epsilon^{-1}$ respectively. A qualitative difference between these two terms is that the fermion hopping $\epsilon$ to the lowest order modifies $J$. Its effect can be seen in a decrease of the period of the oscillations at $\epsilon = 0.1$. Beyond this point, there are only quantitative differences between the two cases, and the time averaged values look similar to the eye.

A different phenomenology is observed in the case of the $z$-Ising coupling $h_z$, shown in Fig.~\ref{fig: Int Breaking}(c). In this case there is little appreciable deviation in the time averaged value for $h_z/J = 0.001,0.01$, beyond the damping of the oscillations. When the coupling is increased to $h_z = 0.1J$, we finally see appreciable decay of the imbalance, but that does not convincingly vanish, compared to the case for other terms. This can be understood by considering the (anti-)commutation relations of $\hat{\sigma}^z_{j,j+1}$ with the charges $\hat{q}_j$, which are
\begin{equation}
\begin{aligned}
\{\hat{\sigma}^z_{j,j+1},\hat{q}_{k}\} &= 0, \quad k= j, j+1,\\
[\hat{\sigma}^z_{j,j+1},\hat{q}_{k}] &= 0, \quad k\neq j, j+1.
\end{aligned}
\end{equation}
We can then make an identification with spin operators $\hat{q}_j \rightarrow \hat{q}_k^z$ and $\hat{\sigma}^z_{j,j+1} \rightarrow \hat{q}_j^x\hat{q}_{j+1}^x$, that is, the transverse field induces an effective Ising coupling between the charges. For the transverse field we are then precisely in the framework of heavy-light mixtures, which are generally believed to become ergodic at long-times. However, the Ising coupling $\hat{\sigma}^z_{j-1,j}\hat{\sigma}^z_{j,j+1}$ maps to the next-nearest neighbour Ising coupling $\hat{q}_{j-1}^x\hat{q}_{j+1}^x$ for the charges. Since the lattice is bipartite, the charges interact separately on these two disconnected sublattices. Importantly, this means that the charges -- and by extension the effective disorder potential -- on neighbouring sites do not become correlated. This leads to increased persistence of localization seen in Fig.~\ref{fig: Int Breaking}(c). 

Since we still have a heavy-light mixture with the addition of the $z$-Ising coupling, we can ask whether this additional persistence survives in the thermodynamic limit at long times, which would be in contrast with the standard phenomenology of these systems. Fig.~\ref{fig: Int Breaking Scaling} shows density imbalance as a function of the system size, which seems to suggest that we also lose localization in this case in the thermodynamic limit, consistent with Ref.~\cite{Papic2015}.

\section{Discussion}\label{sec: discussion}

In this paper we have studied a family of unconstrained $\mathbb{Z}_2$ lattice gauge theories of spinless fermions coupled to spins-1/2. Using these models, we have revealed a general phenomenology of disorder-free mechanism for localization that extends beyond the models we discussed here. The key feature of this mechanism is the presence of an extensive set of local conserved quantities that emerge from the coupling between the spins and the fermions, with fermions attached to defects in the lattice gauge theory. In particular we have outlined the connection to the toric code when our model is defined on a periodic square lattice and the resulting constrained dynamics of defects therein. The coupling to fermions allows us to change the behaviour of defects depending on the fermion configuration and filling.

As a diagnostic of localization, we have considered the dynamics of the system after a global quantum quench, which revealed the persistence of memory of initial states in both one and two-dimensional case. The results for the density of states and for the localization length confirmed the localization picture. Through these numerical experiments and the exact duality mapping of the Hamiltonian and states, we have demonstrated that a disorder-free mechanism that we found in 1D in Refs.~\cite{Smith2017,Smith2017_2} applies much more generally.

The effective disorder that appears in our models has a binary nature in contrast to the continuously sampled disorder, that is usually studied in the context of localization physics. While for small effective disorder strengths the localization picture for binary disorder and uniform disorder should agree qualitatively, in the strong-disorder limit, we reveal new features that are specific to binary disorder. As most clearly demonstrated in the 1D case, for strong disorder, we find that the single-particle spectrum splits up into a discrete set of degenerate energy levels, of which only very few carry the majority of the spectral weight. This gives rise to resonances which result in the observed area-law plateau in the entanglement entropy, and in long-lived fluctuations of the fermion density. These effects are somewhat washed-out with increased dimensionality. Interestingly, in the 2D case, the strong-disorder limit can be understood as a quantum site percolation problem. By biasing the distribution of the conserved charges above the percolation threshold, we found a delocalization transition of fermions by the percolation mechanism. This behaviour was demonstrated by studying a quantum quench from a domain wall initial state, the density of states, and the localization length, which confirmed a direct correspondence with the threshold for classical site percolation. Importantly, our delocalization transition does not require spatial correlations in the disorder, compared to previously known cases. Due to the lower percolation threshold in 3D, for the cubic lattice we expect that one can observe two delocalized phases for small and large effective disorder $h/J$, separated by an intermediate localized phase.

While the model we consider in Eq.~\eqref{eq: H general} can be mapped to a model of free fermions, we also considered effects of terms which render the model truly interacting in one dimension. These terms fall into two categories -- those that commute with the charges and those that don't. In the first category we studied nearest-neighbour density interactions, and a longitudinal field. In both cases, in certain parameter regimes, we observed the logarithmic entanglement growth characteristic of many body localization. Furthermore, for the longitudinal field, we also observed sub-logarithmic growth, which shows a good fit with $\ln(\ln(t))$ behaviour.

The second class of integrability-breaking terms are those which give dynamics to the conserved charges. These terms generate dynamics similar to that studied in heavy-light mixtures. The latter showed quasi-MBL behaviour, i.e., only transient localization. We indeed observe quasi-MBL in the region of parameters which we study. However, we find a special case that was comparatively less effective at destroying the localization behaviour. This was due to the dynamics being along the two disconnected chains in the bipartite lattice, which means that nearest-neighbour correlations weren't generated. Despite this, it also appears to lead to ergodic behaviour in the thermodynamic limit at long-times.

The models that we discuss in this paper are particular relevant and timely because of the recent experimental progress in the control of isolated quantum systems in cold atoms experiments~\cite{Choi2016,Schreiber2015}. Similar to the recent simulation of the Schwinger model~\cite{Martinez2016}, our models can be implemented with current technological capabilities. Furthermore, they provide minimal models for lattice gauge theories where in the non-interacting case we are able to perform large scale numerical simulations. Therefore they can serve as a benchmark about which we can consider the truly interacting perturbations which are practically impossible to simulate in anything other than finite 1D chains.

\section*{Acknowledgements} We are grateful to C. Castelnovo for enlightening discussions. A.S.~would like to acknowledge the EPSRC for studentship funding under Grant No.~EP/M508007/1. R.M.\ was in part supported by DFG under grant SFB 1143. The work of  D.K.~was supported by EPSRC Grant No.~EP/M007928/2. \\

\appendix

\noindent\makebox[\linewidth]{\resizebox{0.7\linewidth}{1pt}{$\bullet$}}\bigskip

\begin{center}
\textbf{Appendices}
\end{center}

In the following Appendices we present details of the numerical methods utilized in our calculations. We also include a discussion of the parameters used to produce the figures shown in the main text. In Appendix~\ref{ap: determinant} we outline the method for calculating fermionic correlators using determinants for the case of bilinear Hamiltonians. We explain how to use the Krylov subspace method for time evolution in Appendix~\ref{ap: Krylov}. In Appendix~\ref{ap: KPM} we present an outline of the kernel polynomial method. And in Appendix~\ref{ap: transfer matrix} we discuss calculations of localization length using transfer matrix techniques.


\section{Calculation of fermion correlators using determinants}\label{ap: determinant}

In the case of a Hamiltonian bilinear in fermion operators, dynamic correlation functions can be obtained in terms of determinants. In the main text we have shown that general correlators for our set of models can be written in terms of purely fermionic correlators. In the following we explain how the calculation of this correlators can be reduced to determinants, see e.g.~\cite{Kovrizhin2010}. A mapping to the free-fermion Hamiltonian dramatically decreases the computational cost compared with ED, which allows us to reach much larger system sizes.

Generically, we are interested in computing expressions of the form
\begin{equation}\label{eq: exp det}
\la \alpha | \exp\{ i \sum_{ij} A_{ij} \hat{c}^\dag_i \hat{c}_j \} | \beta \ra,
\end{equation}
where $A$ is a Hermitian matrix, and $|\alpha \ra = \hat{c}^\dag_{m_N} \cdots \hat{c}^\dag_{m_1} | vac \ra$ and $|\beta \ra = \hat{c}^\dag_{n_N} \cdots \hat{c}^\dag_{n_1} | vac \ra$ are fermionic Slater determinants. To proceed with the calculation of (\ref{eq: exp det}) we first use the unitarity of the exponential operator $\hat{U}_A \equiv \exp\{ i \sum_{ij} A_{ij} \hat{c}^\dag_i \hat{c}_j \}$ to rewrite (\ref{eq: exp det}) as
\begin{multline}\label{eq: c tilde}
\la vac | \hat{c}_{m_1} \cdots \hat{c}_{m_N} \hat{U}_A \hat{c}^\dag_{n_N}\hat{U}_A^\dag \cdots \hat{U}_A\hat{c}^\dag_{n_1}\hat{U}_A^\dag \hat{U}_A|vac \ra \\
= \la vac | \hat{c}_{m_1} \cdots \hat{c}_{m_N} \hat{\tilde{c}}^\dag_{n_N} \cdots \hat{\tilde{c}}^\dag_{n_1} |vac \ra,
\end{multline}
where $\hat{\tilde{c}}^\dag_j \equiv \hat{U}_A \hat{c}^\dag_{j}\hat{U}_A^\dag$ and we use $\hat{U}_A |vac\ra = |vac\ra$. With the help of the Baker-Hausdorff formula we obtain
\begin{equation}\label{c_tilde}
\hat{\tilde{c}}^\dag_i = \sum_j\exp\{i A^T\}_{ij} \hat{c}^\dag_j \equiv \sum_j U^T_{A,ij} \hat{c}^\dag_j,
\end{equation}
distinguishing between the operator $\hat{U}_A$, and the matrix $U_A$ by a hat. Finally, we insert (\ref{c_tilde}) into~\eqref{eq: c tilde}, and use the fermionic anti-commutation relations to obtain
\begin{equation}\label{eq: det expression}
\la \alpha | \hat{U}_A |\beta \ra = \det D, \quad D_{jk} = [U_A]_{n_j m_k},
\end{equation}
with $j,k = 1,\dots N$. In other words we select from the matrix $U_A$ those lines and columns that correspond to occupied states in Slater determinants $|\beta\rangle,|\alpha\rangle$.

This derivation allows for further generalisations. For example, in the case of an arbitrary number of unitary operators, using repeatedly the Baker-Hausdorff formula, we obtain
\begin{equation}\label{eq:prod_U}
\la \alpha | \hat{U}_A \hat{U}_B \cdots |\beta \ra = \det D, \quad D_{jk} = [U_A U_B \cdots]_{n_j m_k}.
\end{equation}
This equation is suitable for evaluating correlators similar to the one in Eq.~\eqref{eq: spin average}. In case of the fermion correlators e.g.~\eqref{eq: exp det} we need to consider expressions of the following form
\begin{equation}
C_{kl} = \la \alpha | \hat{c}^\dag_k \exp\{i \sum_{ij} A_{ij}\hat{c}^\dag_i \hat{c}_j \} \hat{c}_l | \beta \ra.
\end{equation}
By commuting $\hat{c}^\dag_k$ to the left, and $\hat{c}_l$ to the right, we pick up factors $(-1)^{N-p}$ and $(-1)^{N-q}$, where $m_p = k$ and $n_q = l$ and arrive at
\begin{equation}
\begin{aligned}
C_{kl} = (-1)^{p+q}\la vac | &\hat{c}_{m_1} \cdots \hat{c}_{m_{p-1}} \hat{c}_{m_{p+1}} \cdots \hat{c}_{m_N}\\
&\times \hat{U}_A \hat{c}^\dag_{n_N} \cdots \hat{c}^\dag_{n_{q+1}} \hat{c}^\dag_{n_{q-1}} \cdots \hat{c}^\dag_{n_1} |vac \ra.
\end{aligned}
\end{equation}
In this case we need to remove the $q$-row and the $p$-column from the matrix $D$ before taking the determinant and then multiply by the corresponding sign. Specifically we need the $q-p$ cofactor of $D$ where $D$ is given in Eq.~\eqref{eq: det expression}. The final expression fo the fermion correlator now can be written in a simple form,
\begin{equation}
C_{kl} = D^{-1}_{lk} \det D,
\end{equation}
where $D_{jk} = [U_A]_{n_j m_k}$, $j,k = 1,\dots N$.

The free-fermion mapping presented in the main text allows one to extract dynamical correlators for system sizes far beyond exact diagonalization. We can estimate the size of the fermionic Hilbert space at half-filling as $N^{-1/2}2^N$ with the spin degrees of freedom adding another factor of $2^N$. Instead of diagonalizing exponentially large matrices the identification of conserved charges allows us to sample uniformly from $\sim 2^N$ determinants of $N\times N$ matrices, corresponding to different charge configurations. Finally, finite-size scaling as well as exact results (up to $N=20$) show that the required number of samples for a given accuracy scales polynomially with $N$. Typically we sample over $10^3-10^4$ charge configurations.

\section{Krylov subspace decomposition}\label{ap: Krylov}

A major bottleneck for computing dynamics using exact diagonalization is the memory requirement for storing many-body wave functions. Although this can be drastically reduced if the model has conserved quantities, memory is still the limiting factor due to the exponential growth of the Hilbert space dimension. A way around this for computing dynamical quantities is to use a smaller set of basis states to perform the time evolution via short time steps. An optimal basis of states for this method can be identified with the Krylov subspace generated by the Hamiltonian. In this Appendix we provide an outline of this method.

Our goal is to efficiently calculate the time evolution of a quantum state $|\Psi(t)\ra = e^{-i\hat{H}t}|\Psi\ra$ for which we use Krylov subspace
\begin{equation}
\mathcal{K}_R = \text{span}\{|\Psi\ra,\hat{H}|\Psi\ra,\hat{H}^2|\Psi\ra, \dots, \hat{H}^{R-1}|\Psi\ra \},
\end{equation}
where $\hat{H}$ is our Hamiltonian, $|\Psi\ra$ is an initial state, and $R$ is the number of states in Krylov subspace (that we choose). The idea is that at short enough times the state $|\Psi(t)\ra$ will be predominantly in this subspace as can be seen from a Taylor expansion of the unitary time evolution
\begin{equation}
|\Psi(t)\ra = \sum_{n=0}^\infty \frac{(-i\hat{H}t)^n}{n!} |\Psi\ra.
\end{equation}
Given a basis for the Krylov subspace $\{|v_i\ra\}$, the best approximation to the unitary time evolution is given by $e^{-i\tilde{H}t}$ with
\begin{equation}
\widetilde{H} = \sum_{ij} |v_i\ra \la v_i| \hat{H} |v_j\ra \la v_j|.
\end{equation}
Therefore, we need to diagonalize a matrix of dimension $R$ with matrix elements $\widetilde{H}_{ij} = \la v_i| \hat{H} |v_j\ra$. For a Hermitian operator the reduced Hamiltonian $\widetilde{H}$ takes a simpler tridiagonal form which can be efficiently diagonalized.

One of the main practical considerations is the accuracy of the Krylov subspace method and the orthogonality between Krylov basis vectors. Due to limited numerical precision the computed Krylov eigenvectors will diverge from the true ones as more of them are included. A way around this is to orthogonalize each new vector to the previous set. However, the numerical errors that arise in this procedure make it also problematic, and eventually orthogonality will be lost. To get around these issues, after roughly 25 applications of matrix multiplications (the number is chosen empirically) we orthonormalize the entire set of vectors using efficient QR decomposition before proceeding further. 

The accuracy of this approximation can be kept below a prescribed threshold only for a finite value of $t$, which is set by the size of the basis dimension $R$. To study time evolution on longer timescales, we use a `restarted evolution method' which computes the time evolution up to a certain time $\delta t$ and then repeats with the new starting state $|\Psi(t+\delta t)\ra$ until the desired time is reached. To check the accuracy of the method we then perform the reverse time evolution and check the difference between the values on the forward and backwards pass, which provides a good estimate of the deviation from the true value~\cite{Sidje1998,Brenes2017}.

The method described in this Appendix is limited by the memory required to store the Hamiltonian $\hat{H}$ and the Krylov basis states. Since the Hamiltonian is typically sparse and has $O(\alpha N)$ non-zero values, the memory requirement scales as $O((R+\alpha)N)$ compared with $O(N^2)$ for exact diagonalization. In our calculations we take $R = 50$, and $\delta t = 1.2$ and compute values for $dt = 0.2$ which we find gives acceptable errors of only 1-2 orders of magnitude above machine precision on the scale of the full time evolution. We note that the computational cost also scales linearly with the number of time steps. 

\section{Density of states: Kernel Polynomial Method}\label{ap: KPM}

The kernel polynomial method (KPM) is a numerical technique for computing spectral quantities. It uses a decomposition into Chebyshev polynomials with modifications to the coefficients to damp Gibbs oscillations which are inherent to Fourier expansion. The benefit of the KPM is that calculations can be reduced to repeated multiplications of the Hamiltonian matrix, which is very efficient for sparse matrices. We will briefly describe the basic procedure, also see Ref.~\cite{Weisse2006} for more details and examples.

\subsubsection*{Chebyshev expansion and modified moments}

Chebyshev polynomials are defined on the interval $[-1,1]$ and form the orthogonal basis with respect to inner products defined on this interval with a special weight function. For Chebyshev polynomials of the first kind $T_n(x)$, this weight function is $w(x) = (\pi \sqrt{1-x^2})^{-1}$, so that the inner product reads,
\begin{equation}
\la f | g \ra_1 = \int^1_{-1} \text{d}x\; \frac{f(x) g(x)}{\pi \sqrt{1-x^2}}, \quad \la T_n | T_m \ra = \frac{1 + \delta_{n,0}}{2} \delta_{n,m}.
\end{equation}
Chebyshev polynomials of the second kind $U_n(x)$ are defined with respect to the weight function $w(x) = \pi \sqrt{1-x^2}$, i.e.
\begin{equation}
\la f | g \ra_1 = \int^1_{-1} \text{d}x\; \pi \sqrt{1-x^2} f(x) g(x), \quad \la U_n | U_m \ra = \frac{\pi^2}{2} \delta_{n,m}.
\end{equation}
These polynomials obey useful recursion relations
\begin{equation}
\begin{aligned}
T_{n+1}(x) &= 2x T_n(x) - T_{n-1}(x),\\
U_{n+1}(x) &= 2x U_n(x) - U_{n-1}(x),
\end{aligned}
\end{equation}
where $T_0(x) = 1$, $T_1(x) = x$, and $U_0(x) = 1$, $U_{-1}(x) = 0$.

Given an orthogonal basis together with the inner product, we can expand a function defined on the interval $[-1,1]$ as 
\begin{equation}
f(x) = \alpha_0 + 2 \sum_{n=1}^\infty \alpha_n T_n(x),
\end{equation}
where the moments are given by
\begin{equation}
\alpha_n = \la f | T_n \ra_1 = \int^1_{-1}\text{d}x\; \frac{f(x)T_n(x)}{\pi\sqrt{1-x^2}}.
\end{equation}
However, the numerical integration of the moments is problematic due to the square root appearing in the denominator. To get around this, we instead define the functions
\begin{equation}
\phi_n(x) = \frac{T_n(x)}{\pi\sqrt{1-x^2}},
\end{equation}
which have the property $\la \phi_n | \phi_m \ra_2 = \la T_n | T_m \ra_1$, and thus we can instead write the expansion as
\begin{equation}\label{eq: Chebyshev expansion}
f(x) = \frac{1}{\pi\sqrt{1-x^2}}\left[\mu_0 + 2 \sum_{n=1}^\infty \mu_n T_n(x)\right],
\end{equation}
where the moments are given by
\begin{equation}
\mu_n = \la f | \phi_n \ra_2 = \int^1_{-1} \text{d}x\; f(x) T_n(x).
\end{equation}

To use this expansion in terms of Chebyshev polynomials we must first rescale the energies and the Hamiltonian so that the bandwidth lies in the interval $[-1,1]$. We thus define the rescaled Hamiltonian and energies
\begin{equation}
\tilde{H} = (H-b)/a, \qquad\qquad \tilde{E} = (E-b)/a,
\end{equation}
where $a=(E_\text{max} - E_\text{min})/(2-\epsilon)$ and $b = (E_\text{max} - E_\text{min})/2$. We include a small factor $\epsilon$ to avoid stability problems near $\pm 1$. In practice we can use analytically obtained bounds on $E_\text{max/min}$ to avoid computing them explicitly. One could also compute $E_\text{max/min}$ for smaller system sizes, add a margin of error and use these for our bounds.

We now have to discretize the function argument and truncate the infinite sum. To make use of the properties of Chebyshev function we choose a set of $K$ points, $x_k = \cos(\pi(k+1/2)/K)$, for $k = 0, \dots, K-1$. With this choice the expansion takes the form
\begin{equation}
f(x_k) = \frac{1}{\pi\sqrt{1-x_k^2}}\left[\mu_0 + 2 \sum_{n=1}^\infty \mu_n \cos\left(\frac{\pi n (k+1/2)}{K}\right)\right].
\end{equation}
Since we are expanding in periodic functions, the truncation of the sums leads to Gibbs oscillations. If we keep the first $M$ terms in the sum, then to remove these oscillations we introduce a kernel of order $M$,
\begin{equation}
K_M(x,y) = g_0 \phi_0(x) \phi_0(y) + 2 \sum_{m=1}^{M-1}g_m \phi_m(x) \phi_m(y),
\end{equation}
which we use to define
\begin{equation}
f_\text{KPM}(x) = \int^1_{-1} \text{d}y\; \pi\sqrt{1-y^2} K_M(x,y) f(y).
\end{equation}
We can then determine the coefficients $g_m$ in the kernel by demanding that $f_\text{KPM}$ is as close as possible to the true function $f(x)$. Closeness can be defined in a number of different ways each of which lead to different set of coefficients. In our calculations we use the Jackson kernel defined by coefficients
\begin{equation}
\begin{aligned}
g_m = \frac{1}{M+1} \Big[ &(M-m+1) \cos\left(\frac{\pi m}{M+1}\right) \\ &+  \sin \left(\frac{\pi m}{M+1}\right) \cot\left(\frac{\pi}{M+1} \right) \Big].
\end{aligned} 
\end{equation}
See Ref.~\cite{Weisse2006} for the derivation of these coefficients and discussions of other choices of kernel. Kernel coefficients are then used to modify the moments in our expansion, and we arrive to an expression
\begin{equation}
\begin{aligned}
f&(x_k) \approx\\ 
&\frac{1}{\pi\sqrt{1-x_k^2}}\left[g_0\mu_0 + 2 \sum_{m=1}^{M-1} g_m\mu_m \cos\left(\frac{\pi m (k+1/2)}{K}\right)\right].
\end{aligned}
\end{equation}

\subsubsection*{Calculation of the moments}

The moments that appear in our expansion are typically of the form $\la \beta | A T_n(H) | \alpha \ra$, where $H$ is the Hamiltonian matrix, $A$ is a matrix representing an operator and $|\alpha\ra$ and $|\beta\ra$ are two states. We need to compute $|\alpha_n \ra \equiv T_n(H) |\alpha\ra$ which can be done using recursion relations
\begin{equation}
|\alpha_{n+1} \ra = 2H |\alpha_n\ra - |\alpha_{n-1} \ra,
\end{equation}
with $|\alpha_0\ra = |\alpha\ra$ and $|\alpha_1\ra = H |\alpha\ra$. If $\beta = \alpha$ and $A = \text{I}$ we can use recursion relations for Chebyshev polynomials, specifically, $2T_m(x) T_n(x) = T_{m+n}(x) + T_{m-n}(x)$, to get
\begin{equation}
\mu_{2n} = 2 \la\alpha_n | \alpha_n \ra - \mu_0, \quad \mu_{2n+1} = \la \alpha_{n+1}| \alpha_n \ra - \mu_1,
\end{equation}
which reduces the number of matrix operations by approximately half.

We also need to compute moments which involve a trace over states. The latter can be computed efficiently using
\begin{equation}
\text{Tr} [AT_n(H)] \approx \frac{1}{R} \sum_{n=1}^{R-1} \la r | A T_n(H) |r \ra,
\end{equation}
where $D$ is the size of the matrix $H$, and $R \ll N$ is the number of chosen random vectors. In all of our calculations we take $R=30$. The random vectors $|r\ra$ are defined through random variables $\varepsilon_{ri}$,
\begin{equation}
|r \ra = \sum_{i = 0}^{N-1}\varepsilon_{ri} |i\ra,
\end{equation}
where $|i\ra$ are the basis vectors with an identity in the i$^{th}$ entry. The random variables must satisfy the relation
\begin{equation}
\la\la \varepsilon_{ri} \ra\ra = 0, \quad \la\la \varepsilon_{ri} \varepsilon_{r'j} \ra\ra = \delta_{rr'} \delta_{ij},
\end{equation}
that is, they are uncorrelated with zero mean, and have unit mean for their absolute value.

\subsubsection*{Density of states}

As an example of the application of the method, let us consider the calculation of the density of states, which is defined as
\begin{equation}
g(E) = \frac{1}{N}\sum_{k=0}^{N-1} \delta(E-E_k).
\end{equation}
The coefficients of the Chebyshev expansion are then given by
\begin{equation}
\begin{aligned}
\mu_n = \int_{-1}^1 dE \; \rho(E) T_n(E)
& = \frac{1}{N} \sum_{k=1}^{N-1} T_n(E_k) \\
& = \frac{1}{N} \sum_{k=1}^{N-1} \la k | T_n(H) |k \ra \\ 
& = \frac{1}{N} \text{Tr} [ T_n(H) ],
\end{aligned}
\end{equation}
which we can compute using the statistical trace and the expectation values as explained above.

In the main text we use the following parameters for the figures:\\
Fig.~\ref{fig: 1D DOS}(a) $\leftarrow N=10^6, M=1500, K = 2M, R=30$;\\
Fig.~\ref{fig: 1D DOS}(b) $\leftarrow N=10^6, M=7500, K = 3M, R=30$;\\
Fig.~\ref{fig: 2D DOS}(a) $\leftarrow N=(10^3)^2, M=1500, K=2M, R=30$;\\
Fig.~\ref{fig: Nhalf bias}(b) $\leftarrow N=(10^3)^2, M=2500, K=2M, R=30$;\\
where $N$ is the number of sites, $M$ is the number of moments included in the expansion, $K$ is the number of discretization points, and $R$ is the number of random states used in the statistical trace.

\section{Localization length: Transfer Matrix}\label{ap: transfer matrix}

The application of transfer matrix approach in the calculations of the localization length proceeds by considering a system which is cut up into slices, with slices connected via  transfer matrices~\cite{Kramer1993}. From these, we can extract the eigenvalues of a limiting matrix that gives the Lyapunov exponents for our system. For instance, consider a 1D chain with the Hamiltonian
\begin{equation}
\hat{H} = -J \sum_j (\hat{c}^\dag_j \hat{c}_{j+1} + \text{H.c}) - \sum_j h_i \hat{c}^\dag_j \hat{c}_j.
\end{equation}
The action of this Hamiltonian on an eigenstate $|\psi\ra = \sum_i \psi_i | i \ra$, where $|i\ra$ is the state localized on site $i$ gives the relation
\begin{equation}
E \psi_i = - J \psi_{i+1} - J \psi_{i-1} -h_i \psi_i,
\end{equation}
where $E$ is the eigenvalue of the state $|\psi\ra$. This equation can be written in a compact form by introducing a transfer matrix
\begin{equation}
\left(\begin{array}{c}\psi_{i+1} \\\psi_i\end{array}\right) =   \left(\begin{array}{cc}-\frac{1}{J}(E+h_i) & -1 \\1 & 0\end{array}\right) \left(\begin{array}{c}\psi_{i} \\\psi_{i-1}\end{array}\right).
\end{equation}
In higher dimensions these equations have to be modified slightly, in particular we get
\begin{equation}
E\psi_i = -J\psi_{i+1} - J\psi_{i-1} - H_\text{perp} \psi_i,
\end{equation}
where $\psi_i$ is now a vector of the values of the wavefunction in the orthogonal direction at the slice $i$, and $H_\text{perp}$ is the matrix representation of the Hamiltonian in the same direction. The transfer matrix equation assumes the form
\begin{equation}
\left(\begin{array}{c}\psi_{i+1} \\\psi_i\end{array}\right) =   \left(\begin{array}{cc}-\frac{1}{J}(E+H_\text{perp}) & -1 \\ 1 & 0\end{array}\right) \left(\begin{array}{c}\psi_{i} \\\psi_{i-1}\end{array}\right).
\end{equation}

Given the transfer matrix, we can compute the product along a long chain of length $L$,
\begin{equation}
Q_L = \prod_{i=1}^L T_i,
\end{equation}
where $T_i$ is the transfer matrix for slice $i$. Oseledec's theorem states that there exists a limiting matrix 
\begin{equation}
\Gamma = \lim_{L\rightarrow\infty} (Q_L Q_L^\dag)^{1/2L},
\end{equation}
with eigenvalues $\exp(\gamma_i)$, where $\gamma_i$ are the Lyapunov exponents of the matrix $Q_L$. The smallest Lyapunov exponent describes the slowest growth of the wavefunction and corresponds to the inverse of the localization length $\lambda$.

More intuitively, we can consider $Q_L$ as the transfer matrix between the extreme ends of the chain:
\begin{equation}
\left(\begin{array}{c}\psi_{L+1} \\\psi_L\end{array}\right) =   Q_L \left(\begin{array}{c}\psi_{2} \\\psi_{1}\end{array}\right),
\end{equation}
so the eigenvalues of $Q_L$ describe the growth of the wavefunction along the length of the system. We can then take the smallest eigenvalue, $q$, of the matrix $Q_L$, and compute the localization length via
\begin{equation}
\lambda = \frac{L}{\ln(q)}.
\end{equation}

The procedure of computing a matrix product of a large number of matrices is numerically unstable since the matrix elements diverge or vanish exponentially. We therefore must orthonormalize the product of matrices after a few number steps. The numerical procedure proceeds as follows: we iteratively construct the product matrix $Q$ by applying randomly generated transfer matrices $T$. After the number of applications exceeds a predefined limit or the amplitude of the elements of the matrix exceed a threshold, we store the logarithm of the eigenvalues and orthonormalize the matrix $Q$, which can be done efficiently using QR-decomposition. We continue applying $T$, storing eigenvalues and orthonormalizing until we have reached the length of the chain $L$. See below a pseudo-code of the algorithm for computing the localization length $\lambda$:

\begin{algorithmic}
	\State $Q \gets Id$
	\State $l \gets 1$
	\While {$l \leq L$}
	\State $count \gets 1$
	\While {$\max(\text{abs}(Q))\leq theshold$ \textbf{or} $count \leq limit$}
	\State Initialise random $T$ for slice
	\State $Q \gets TQ$
	\State $l$++
	\State $count$++
	\EndWhile
	\State $b \gets \text{eig}(Q)$
	\State $c \gets c + \ln(b)$
	\State $Q \gets \text{orthonormal}(Q)$
	\EndWhile
	\State $b \gets \text{eig}(Q)$
	\State $c \gets c + \ln(b)$
	\State $lambda \gets \max(L/c)$
\end{algorithmic}

\end{document}